\documentclass[twocolumn,tighten]{aastex6}
\usepackage{amsmath,amstext}
\usepackage[T1]{fontenc}

\usepackage{apjfonts}
\usepackage{appendix}
\usepackage[figure,figure*]{hypcap}
\usepackage{listings}
\usepackage{float}
\usepackage{subfigure}
\shorttitle{Orbital decay of hot Jupiters}
\shortauthors{Patra et al.}

\begin{document}

\title{The continuing search for evidence of tidal orbital decay of
  hot Jupiters\footnote{This paper includes data gathered with the
    6.5\,m Magellan/Clay Telescope located at Las Campanas
    Observatory, Chile.}}

\author{Kishore C.~Patra\altaffilmark{1}}
\author{Joshua N.~Winn\altaffilmark{2}}
\author{Matthew J.~Holman\altaffilmark{3}}
\author{Michael~Gillon\altaffilmark{6}}
\author{Artem~Burdanov\altaffilmark{6}}
\author{Emmanuel~Jehin\altaffilmark{7}}
\author{Laetitia~Delrez\altaffilmark{5,6}}
\author{Francisco J.~Pozuelos \altaffilmark{6,7}}
\author{Khalid ~Barkaoui \altaffilmark{6,13}}
\author{Zouhair ~Benkhaldoun \altaffilmark{13}}
\author{Norio~Narita\altaffilmark{8,9,10,11}}
\author{Akihiko~Fukui\altaffilmark{11,12}}
\author{Nobuhiko~Kusakabe\altaffilmark{8}}
\author{Kiyoe~Kawauchi\altaffilmark{12}}
\author{Yuka~Terada\altaffilmark{14}}
\author{L.G.~Bouma\altaffilmark{2}}
\author{Nevin N.~Weinberg\altaffilmark{4}}
\author{Madelyn~Broome\altaffilmark{2}}

\altaffiltext{1}{Department of Astronomy, University of California, Berkeley, CA 94720, USA}

\altaffiltext{2}{Department of Astrophysical Sciences, Princeton
  University, 4 Ivy Lane, Princeton, NJ 08544, USA}

\altaffiltext{3}{Harvard-Smithsonian Center for Astrophysics, 60
  Garden Street, Cambridge, MA 02138, USA}

\altaffiltext{4}{Department of Physics, and Kavli Institute for
  Astrophysics and Space Research, Massachusetts Institute of
  Technology, Cambridge, MA 02139, USA}

\altaffiltext{5}{Cavendish Laboratory, J.J.\ Thomson Avenue, Cambridge, CB3 0HE, UK}

\altaffiltext{6}{Astrobiology Research Unit, Universit\'e de Li\`ege,
  All\'ee du 6 ao\^ut 17, Sart Tilman, 4000, Li\`ege 1, Belgium}

\altaffiltext{7}{STAR Research Unit, Universit\'e de Li\`ege, All\'ee
  du 6 ao\^ut 17, Sart Tilman, 4000, Li\`ege 1, Belgium}

\altaffiltext{8}{Astrobiology Center, 2-21-1 Osawa, Mitaka, Tokyo 181-8588, Japan}

\altaffiltext{9}{JST, PRESTO, 2-21-1 Osawa, Mitaka, Tokyo 181-8588, Japan}

\altaffiltext{10}{National Astronomical Observatory of Japan, 2-21-1 Osawa, Mitaka, Tokyo 181-8588, Japan}

\altaffiltext{11}{Instituto de Astrof\'isica de Canarias, V\'ia L\'actea s/n, E-38205 La Laguna, Tenerife, Spain}

\altaffiltext{12}{Department of Earth and Planetary Science, Graduate
  School of Science, The University of Tokyo, 7-3-1 Hongo, Bunkyo-ku,
  Tokyo 113-0033, Japan}

\altaffiltext{13}{Oukaimeden Observatory, High Energy Physics and Astrophysics Laboratory, Cadi Ayyad
University, Marrakech, Morocco}

\altaffiltext{14}{Department of Astronomy, Graduate School of Science, The University of Tokyo, 7-3-1 Hongo, Bunkyo-ku, Tokyo 113-0033, Japan}

\begin{abstract}

  Many of the known hot Jupiters are formally unstable to tidal
  orbital decay.  The only hot Jupiter for which orbital decay has
  been directly detected is WASP-12, for which transit timing
  measurements spanning more than a decade have revealed that the
  orbital period is decreasing at a rate of $dP/dt\approx 10^{-9}$,
  corresponding to a reduced tidal quality factor of about $2\times
  10^5$.  Here, we present a compilation of transit-timing data for
  WASP-12 and eleven other systems which are especially favorable for
  detecting orbital decay: KELT-16; WASP-18, 19, 43, 72, 103, 114, and
  122; HAT-P-23; HATS-18; and OGLE-TR-56. For most of these systems we
  present new data that extend the time baseline over which
  observations have been performed.  None of the systems besides
  WASP-12 displays convincing evidence for period changes, with
  typical upper limits on $dP/dt$ on the order of $10^{-9}$ or
  $10^{-10}$, and lower limits on the reduced tidal quality factor on
  the order of $10^5$.  One possible exception is WASP-19, which shows
  a statistically significant trend, although it may be a spurious
  effect of starspot activity. Further observations are encouraged.

\end{abstract}

\keywords{Exoplanets --- transits --- exoplanet tides}

\section{Introduction}
\label{sec:intro}

The discovery of 51~Pegasi~b, the first known hot Jupiter, was
announced nearly a quarter of a century ago \citep{MayorQueloz1995}.
Now, there are several hundred confirmed hot Jupiters listed in the
NASA Exoplanet
Archive.\footnote{\url{https://exoplanetarchive.ipac.caltech.edu/}}
Many questions remain unanswered about how these short-period gas
giants formed and came to occupy orbits so close to their host stars
\citep{DawsonJohnson2018}.

The future of these planets is also uncertain. Theoretically, it is
straightforward to show that the orbits of many of the known hot
Jupiters should be slowly shrinking, due to tidal interactions between
the star and planet, and the accompanying long-term dissipation of
energy and transfer of angular momentum from the orbit to the star
\citep{Rasio+1996, Sasselov2003, Levrard+2009}.  Tidal evolution
should eventually lead to planet engulfment if the orbital angular
momentum is too low to allow the star to achieve spin-orbit synchronization 
\citep{Hut1980}. However, the timescale of the process is
unknown and difficult to calculate theoretically, because of our
incomplete understanding of the physical processes by which the energy
of tidal oscillations is ultimately converted into heat
\citep{Ogilvie2014}. The dissipation rate depends on the mass and
orbital period of the planet, and probably also on the interior
structure of the star, e.g., whether the star has a thick convective
envelope, or a convective core \citep{BarkerOgilvie2010,
  EssickWeinberg2016, Weinberg2017}.

Population studies have provided some evidence that orbital decay does
occur on astrophysically relevant timescales. Some examples are the
scarcity of gas giants with periods less than a day \citep[see,
  e.g.][]{Jackson+2008,Hansen2010,Penev+2012,Ogilvie2014}, the
anomalously rapid rotation of some hot-Jupiter host stars
\citep{Penev+2018} and the rarity of hot Jupiters around subgiant
stars \citep{VillaverLivio2009,Hansen2010,SchlaufmanWinn2013}.  Tidal
decay might also be responsible for the lower occurrence of close-in
planets around rapidly rotating stars \citep{TeitlerKonigl2014}, or
the realignment of stars and their planetary orbits
\citep{MatsakosKonigl2015}.

The goal of the work described in this paper is to seek more direct
evidence for tidal orbital decay through long-term timing of the
transits of hot Jupiters. These types of observations have already led
to the detection of an apparent period decrease in the WASP-12 system
\citep{Maciejewski+2016,Patra+2017,Yee+2020}. Another candidate that
recently emerged is WASP-4 \citep{Bouma+2019}, although in that case,
the evidence is not as compelling.

This paper is organized as follows. Section~\ref{sec:targets}
describes our compilation of other promising candidates for detecting
orbital decay. Section \ref{sec:newtimes} presents new transit-timing
data for many of these systems, and Section~\ref{sec:timing} presents
analyses of all the available data. Section~\ref{sec:timing} also
revisits the WASP-12 system, to address a nagging concern that the
apparent timing anomalies are actually systematic errors due to
unmodeled starspots affecting the transit light curve. Since this type
of systematic error would likely be chromatic, we undertook multicolor
photometry of several transits. Section~\ref{sec:summary}
summarizes the results.

\section{Target Selection}
\label{sec:targets}

For the simple and traditional model of tidal interactions in which
the equilibrium tidal bulge is tilted away from the line
joining the planet and star by a small and constant angle,
the rate of change of the orbital period is
\begin{equation}
  \label{eq:dpdt}
    \frac{dP}{dt} = - \frac{27\pi}{2Q'_{\star}} \left( \frac{M_{\rm p}}{M_{\star}}\right) \left(\frac{R_{\star}}{a} \right)^{5},
\end{equation}
where $Q'_\star$ is the ``reduced tidal quality factor,'' a
dimensionless number\footnote{Specifically, $Q'_\star \equiv 1.5
  Q_\star / k_{2,\star}$, where $Q_\star$ is the tidal quality factor,
  and $k_{2,\star}$ is the tidal Love number \citep{Ogilvie2014}.}
quantifying the stellar tidal dissipation rate; $M_{\rm p}$ and
$M_\star$ are the planetary and stellar masses; $R_\star$ is the
stellar radius; and $a$ is the orbital radius. This expression is
based on the assumptions that the orbit is circular, the star's
angular velocity is much smaller than the orbital angular velocity,
and the tidal dissipation within the planet can be neglected.  It can
be obtained by applying Kepler's third law to Equation (20) of
\citet{GoldreichSoter1966} or Equation (12) of \citet{Ogilvie2014}.

\begin{figure*}
	\centering
	\includegraphics[width=0.9\linewidth]{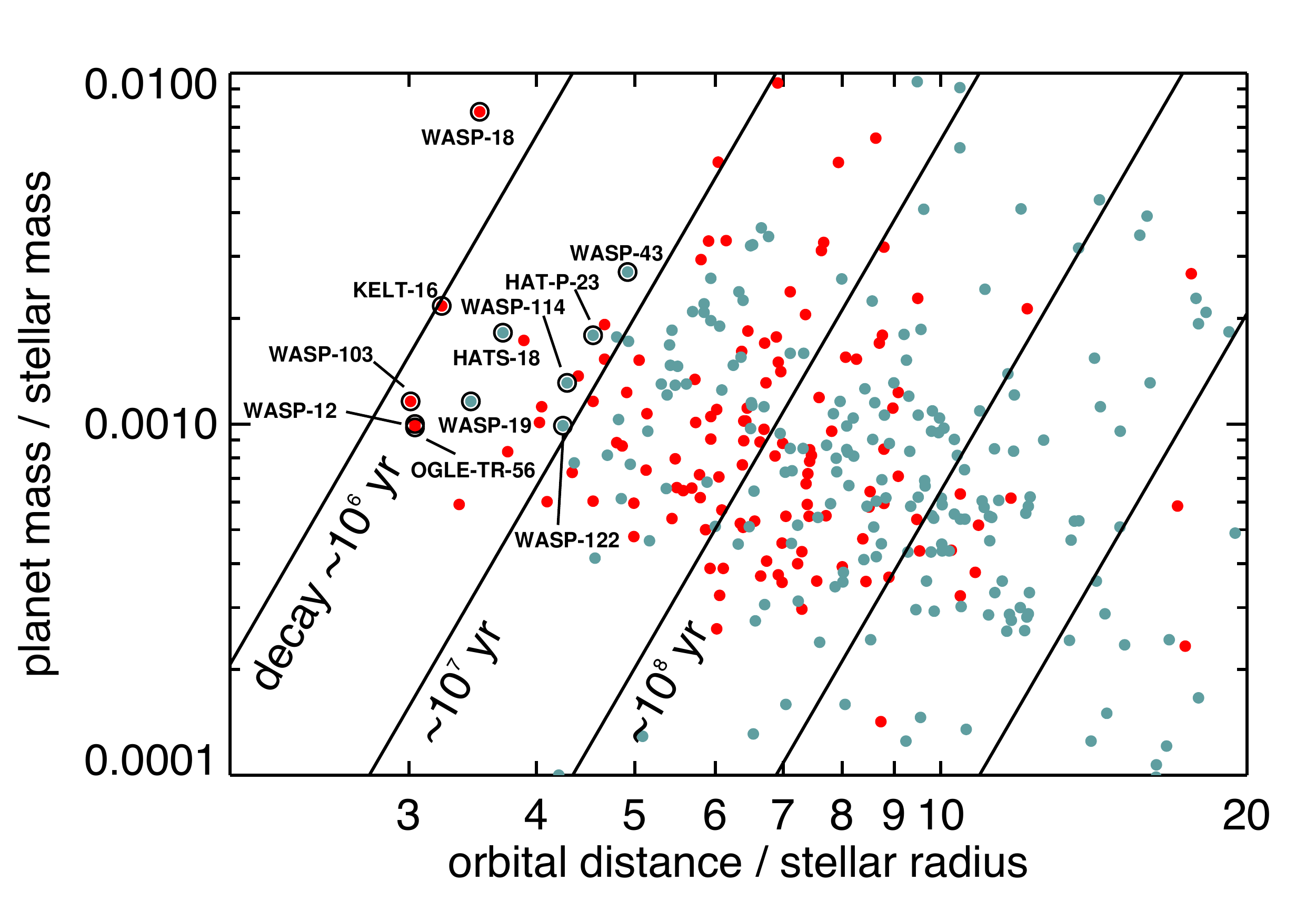}
	\caption{Key parameters for determining the rate of orbital
          decay.  Each point represents one of the known transiting
          planets, with red symbols for those with ``hot'' host stars
          ($T_{\rm eff} > 6000$\,K) and blue symbols for those with
          cooler host stars.  The contours of constant decay timescale
          are based on Equation~(\ref{eq:dpdt}) using a nominal value
          $Q_\star'=10^6$. The most favorable candidates are circled
          and labeled. }
	\label{fig:candidates}
\end{figure*}

This theory is undoubtedly oversimplified, and the dissipation by
dynamical tides may be more important than the equilibrium
tides. Nevertheless, the dependence on the mass ratio and the strong
dependence on orbital separation are generic features of tidal
theories. Therefore, to select good candidates for detecting orbital
decay, we ranked the known transiting planets according to the value
of $(M_{\rm p}/M_\star)(R_\star/a)^5$, which is proportional to
$dP/dt$ according to Equation~(\ref{eq:dpdt}).  We restricted the
sample to host stars brighter than $V=14.5$ to make observations
practical with meter-class telescopes, although we made an exception
for OGLE-TR-56 ($V=16.6$) because of the long time interval over which
data are available.  We also made a distinction between ``hot'' and
``cool'' stars, with the dividing line at $T_{\rm eff}=6000$\,K. This
is because tidal dissipation is expected to be more rapid in cool
stars, due to their thicker outer convective envelopes --- although it
is worth noting that the rapidly decaying hot Jupiter WASP-12 has a
host star with $T_{\rm eff} > 6000$\,K \citep{Hebb+2009, Torres+2012,
  Mortier+2013}.  Figure~\ref{fig:candidates} displays the values of
the key parameters $M_{\rm p}/M_\star$ and $a/R_\star$ for the known
transiting planets, with hot and cool stars depicted in different
colors.  Some contours of constant $P/\dot{P}$ are labeled with the
predicted decay timescale according to Equation~\ref{eq:dpdt},
assuming a nominal value of $Q_\star'=10^6$.

We decided to focus on a dozen objects: the 5 top-ranked hot stars,
and the 7 top-ranked cool stars. They are circled and identified in
Figure~\ref{fig:candidates}, and listed in Table~\ref{tbl:targets}.
The list includes some stars that have been observed for a decade, for
which we might hope to detect orbital decay in the near term, and
others that were more recently discovered, for which we wanted to
establish anchor points for future monitoring.  We limited the sample
to a dozen objects simply to keep the project manageable in
scope. There are other systems that are very nearly as good that we do
not describe in this paper, but that are also worth further attenton
in the future: HATS-70b, WASP-167b, CoRoT-14b, WASP-173b, WASP-87b and
HAT-P-7b, to name a few.

%\begin{rotatetable}
\begin{table*}
	%\small
	%\centering
	
\resizebox{\textwidth}{!}{
\hskip-2.5cm\begin{tabular}{lcccccccccccl}
	\hline 
	\hline
	Name & $M_\star/M_\odot$ & $R_\star/R_\odot$ & $T_{\rm eff}$ & $M_{\rm p}/M_{\rm Jup}$ &
               $R_{\rm p}/R_{\rm Jup}$ & $a/R_\star$ & $P$ & Age & $v\sin i_\star$ & N\textsubscript{obs} & Figure of & Ref.\\ 
        & & & [K] & & & & [d] & [Gyr] & [km s\textsuperscript{-1}]& & merit &  \\
	\hline
	WASP-18    & 1.46(29)  & 1.29(05)  & 6431(48)  & 11.4(1.5) & 1.20(05) & 3.562(23)  & 0.94 & 1.0(0.5) & 11.0(1.5) &  1 & 16.6 & 1, 2, 3  \\
        % Ms, Rs, Teff, Mp, Rp from Stassun+2017; a/R from Shporer+2019; age and vsini from Hellier+2009
	KELT-16    & 1.211(46) & 1.360(64) & 6236(54)  & 2.75(16)  & 1.415(84) & 3.23(13)  & 0.97 & 3.1(0.3) & 7.6(0.5)  &  2 & 8.6  & 4 \\
        % Teff, Oberst+2017
        WASP-103   & 1.21(11)  & 1.416(43) & 6110(160) & 1.51(11)  & 1.623(53) & 3.010(13) & 0.93 & 4(1)     & 10.6(0.9) &  4 & 6.7  & 5, 6 \\
        % Delrez+2018 for everything except vsini which is from Gillon+2014
	WASP-12    & 1.434(11) & 1.657(46) & 6360(140) & 1.470(76) & 1.900(57) & 3.039(34) & 1.09 & 2(1)     & $<\,2.2$  &  6 & 5.3  & 7, 8 \\
        % everything from Collins+2017 except age and vsini limit which are from  Hebb+2009
	HATS-18    & 1.037(47) & 1.020(57) & 5600(120) & 1.980(77) & 1.34(10)  & 3.71(22)  & 0.84 & 4.2(2.2) & 6.23(47)  &  2 & 3.6  & 9 \\
        % \citet{Penev+2016}
	WASP-19    & 0.935(41) & 1.018(15) & 5460(90)  & 1.139(36) & 1.410(21) & 3.45(14)  & 0.79 & 10.2(3.8)& 4(2)      &  3 & 3.3  & 10, 11 \\
        % masses/radii/age/a/R from \citet{Mancini+2013}, vsini from Hebb+2010
	OGLE-TR-56 & 1.228(78) & 1.363(89) & 6050(100) & 1.39(18)  & 1.363(92) & 3.74(19)  & 1.21 & 3.1(1.2) & 3(1)      &  2 & 2.1  & 12, 13 \\
        % everything from Torres+2008 except rotation which is from Sasselov2003
       	HAT-P-23   & 1.130(35) & 1.203(74) & 5905(80)  & 2.09(11)  & 1.368(90) & 4.14(23)  & 1.21 & 4(1)     & 8.1(0.5)  &  3 & 2.0  & 14 \\
        % everything from Bakos+2011.
	WASP-72    & 1.386(55) &  1.98(24) & 6250(100) & 1.546(59) & 1.27(20)  & 4.02(49)  & 2.22 & 3.2(0.6) & 6.0(0.7)  &  2 & 1.4  & 15 \\
        % all from Gillon+2013
	WASP-43    & 0.717(25) & 0.667(11) & 4520(120) & 2.034(52) & 1.036(19) & 4.918(53) & 0.81 & \nodata  & 2(1)      &  3 & 1.3  & 16 \\
        % all from Gillon+2012 except vsini which is from Hellier+2011
	WASP-114   & 1.289(53) & 1.43(60)  & 5940(140) & 1.769(64) & 1.339(64) & 4.29(19)  & 1.55 & 4(2)     & 6.4(0.7)  &  1 & 1.3  & 17 \\
        % all from Barros+2016
	WASP-122   & 1.239(39) & 1.52(03)  & 5720(130) & 1.284(32) & 1.743(47) & 4.248(72) & 1.71 & 5.11(80) & 3.3(0.8)  &  0 & 1.0  & 18 \\
        % all from Turner+2016
	\hline 
\end{tabular}} 
\caption{Favorable targets for the search for tidal orbital decay.
  Parentheses enclose the uncertainties in the last few digits, e.g., 4(1) means $4\pm 1$ and 1.363(92) means $1.363\pm 0.092$.
  N\textsubscript{obs} is the number of new light curves reported in this paper.
  The figure of merit in the second-to-last column is proportional to $(M_{\rm p}/M_\star)(R_\star/a)^5$,
  and normalized to unity for WASP-122.
  References:
  1 - \citet{Stassun+2017};
  2 - \citet{Shporer+2019};
  3 - \citet{Hellier+2009};
  4 - \citet{Oberst+2017};
  5 - \citet{Delrez+2018};
  6 - \citet{Gillon+2014};
  7 - \citet{Collins+2017};
  8 - \citet{Hebb+2009};
  9 - \citet{Penev+2016};
  10 - \citet{Mancini+2013}
  11 - \citet{Hebb+2010};
  12 - \citet{Torres+2008};
  13 - \citet{Sasselov2003};
  14 - \citet{Bakos+2011};
  15 - \citet{Gillon+2012};
  16 - \citet{Hellier+2011};
  17 - \citet{Barros+2016};
  18 - \citet{Turner+2016}.
  Notes:
  For WASP-18, $a/R_\star$ is from Ref.\ 2, while age and $v\sin i_\star$ are from Ref.\ 3.
  For WASP-103, $v\sin i_\star$ is from Ref.\ 6.
  For WASP-12, age and $v\sin i_\star$ are from Ref.\ 8.
  For WASP-19, $v\sin i_\star$ is from Ref.\ 11.
  For OGLE-TR-56, $v\sin i$ is from Ref.\ 13.
}
\label{tbl:targets}

\end{table*}
%\end{rotatetable}

\section{Observations of New transits}
\label{sec:newtimes}

Transits of KELT-16b, WASP-103b, HAT-P-23b, and WASP-114b were
observed with the 1.2m telescope at the Fred Lawrence Whipple
Observatory (FLWO) on Mount Hopkins in Arizona. Time-series photometry
was performed using images from the KeplerCam detector and a Sloan
$r^{\prime}$-band filter. The field of view of this camera is
$23.1$~arcmin on a side. We used  $2\times 2$ binning, giving a
pixel scale of $0.68$\,arcsec. Transits of WASP-19b, WASP-43b, WASP-18b, 
WASP-72b and HATS-18 were observed with the 0.6m TRAPPIST-South 
telescope at Observatorio La Silla, Chile \citep{Trappist1,Trappist2}. 
TRAPPIST-South is equipped with a 2048$\times$2048 pixels FLI Proline PL3041-BB CCD 
camera with a field of view of $22$\,arcmin on a side, giving a pixel 
scale of $0.64$\,arcsec. Two additional transits of WASP-43b were obtained with the TRAPPIST-North telescope at the Oukaimeden Observatory, Morocco \citep{Barkaoui+2019}. 
It is a twin telescope in the Northern hemisphere equipped with a 2000$\times$2000 deep-depletion 
Andor IKONL BEX2 DD CCD camera with a pixel scale of $0.60$\,arcsec and an on-sky 
field of view of $20$\,arcmin. Transits of OGLE-TR-56b were observed
with the 6.5m Magellan Clay Telescope at Las Campanas Observatory in
Chile. Multi-color transits of WASP-12b were observed by the MuSCAT
camera on the 1.88 m telescope at the Okayama Astrophysical
Observatory in Japan \citep{Narita+2015}.

For transits observed before March 2018, raw images were processed by
performing standard overscan correction, debiasing, and flat-fielding
with IRAF.\footnote{The Image Reduction and Analysis Facility (IRAF)
  is distributed by the National Optical Astronomy Observatory, which
  is operated by the Association of Universities for Research in
  Astronomy under a cooperative agreement with the National Science
  Foundation.} For transits observed thereafter, we used AstroImageJ
[AIJ; \citet{Collins+2017}] to perform these same steps.  The
observation time-stamps were placed on the BJD\textsubscript{TDB} time
system using the time utilities code of \citet{Eastman2010}. Aperture
photometry was performed for each target and an ensemble of about 8
comparison stars of similar brightness. The reference signal was
generated by summing the flux of the comparison stars. The flux of the
target star was then divided by this reference signal to produce a
time series of relative flux. This procedure was performed for many
different choices of the aperture radius, and the final radius was
selected to give the smallest possible scatter in the relative flux of
the target star, outside of the transits.

\begin{figure}
	\centering
	\includegraphics[width=1.0\linewidth]{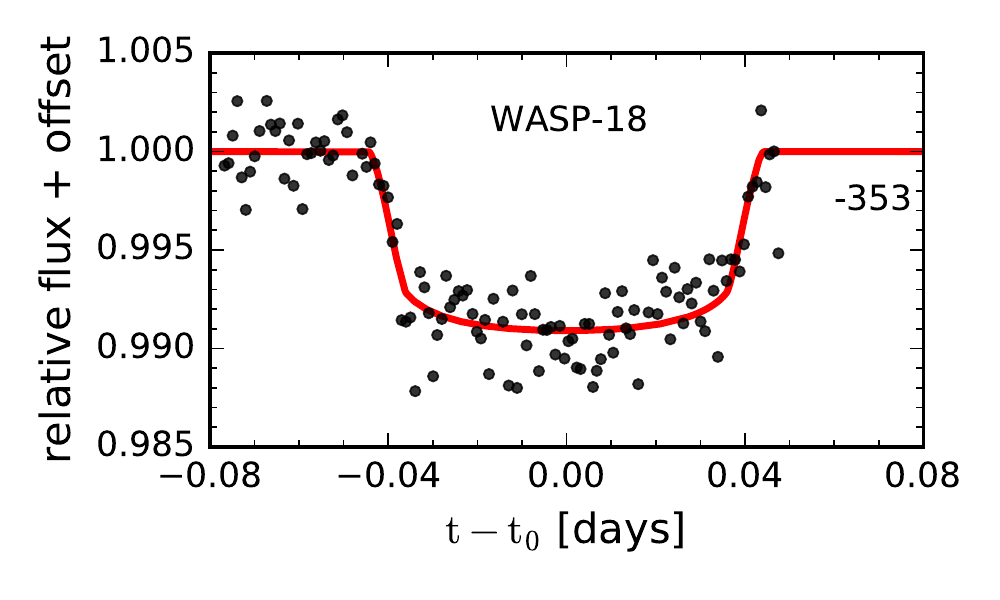}
	\caption{New transit light curve for WASP-18 b. Epoch number is printed to the bottom right of the frame.}
	\label{fig:wasp18lights}
\end{figure}

\begin{figure}
	\centering
	\includegraphics[width=1.0\linewidth]{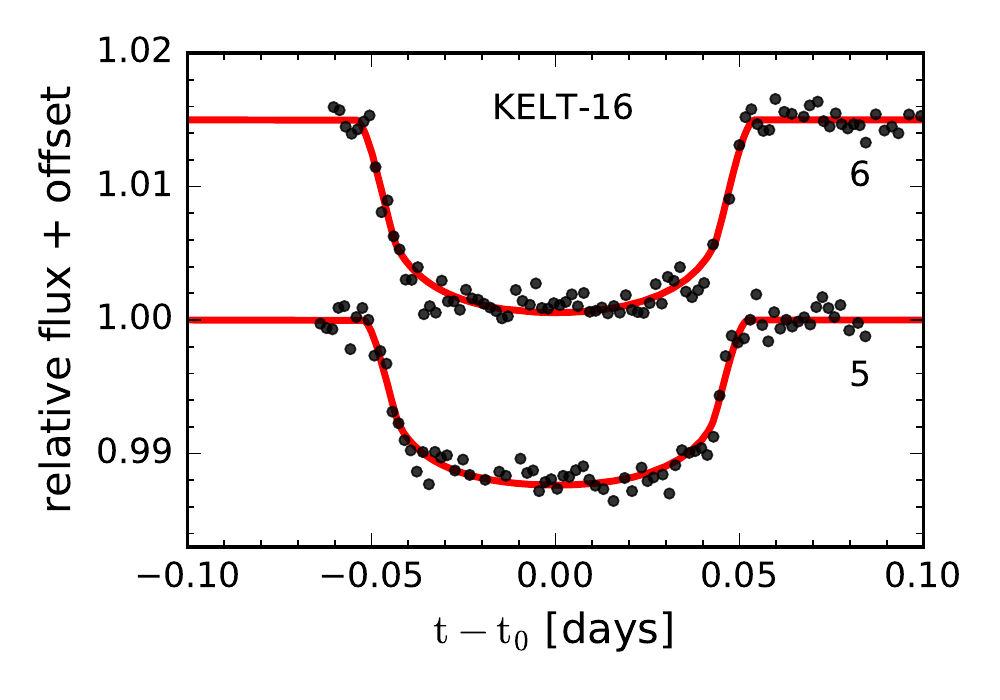}
	\caption{New transit light curves for KELT-16 b. Epoch numbers are printed to the right of each curve. Vertical offsets have been applied to separate the light curves.}
	\label{fig:kelt16lights}
\end{figure}

\begin{figure}
	\centering
	\includegraphics[width=1.0\linewidth]{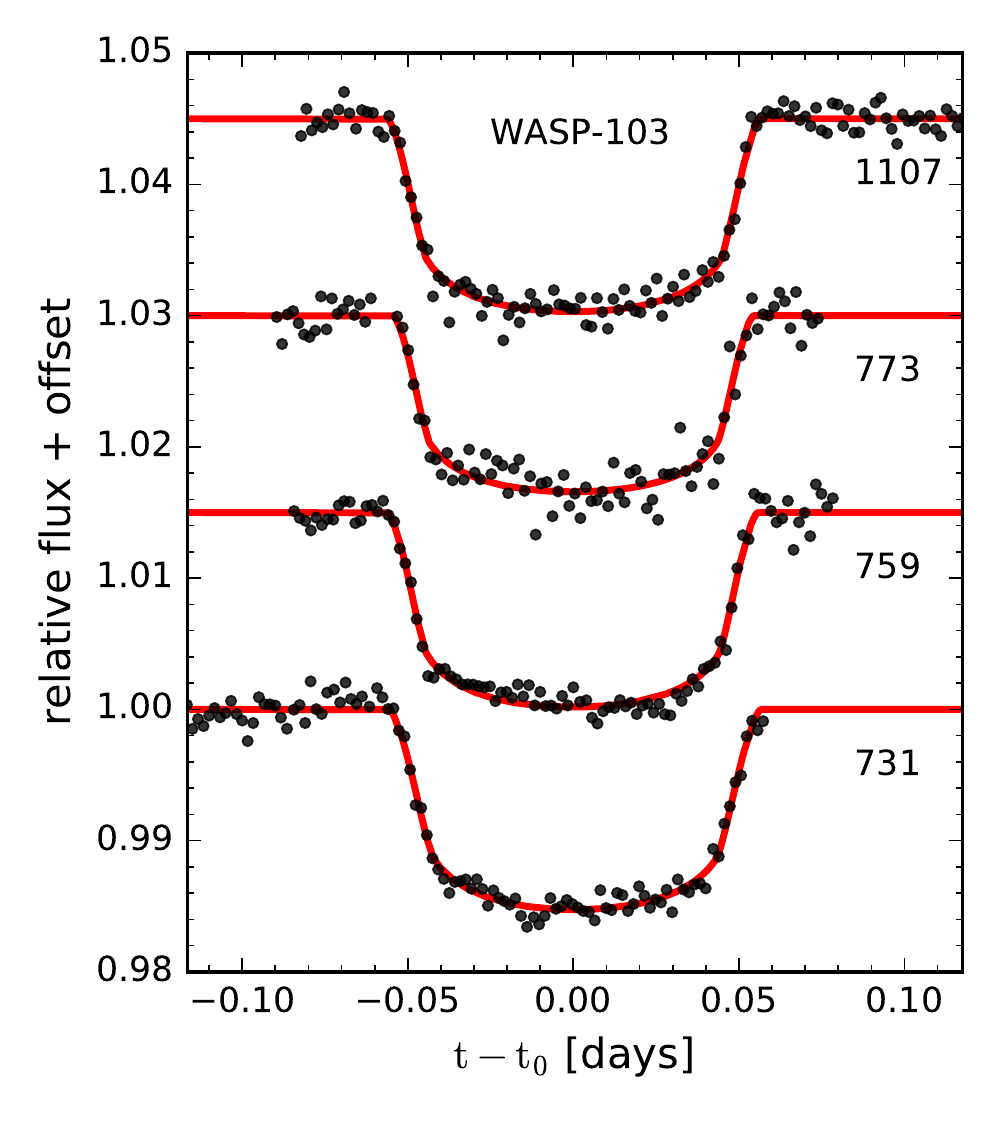}
	\caption{New transit light curves for WASP-103 b. Epoch numbers are printed to the right of each curve. Vertical offsets have been applied to separate the light curves.}
	\label{fig:wasp103lights}
\end{figure}

\begin{figure}[h]
	\centering
	\includegraphics[width=0.95\linewidth]{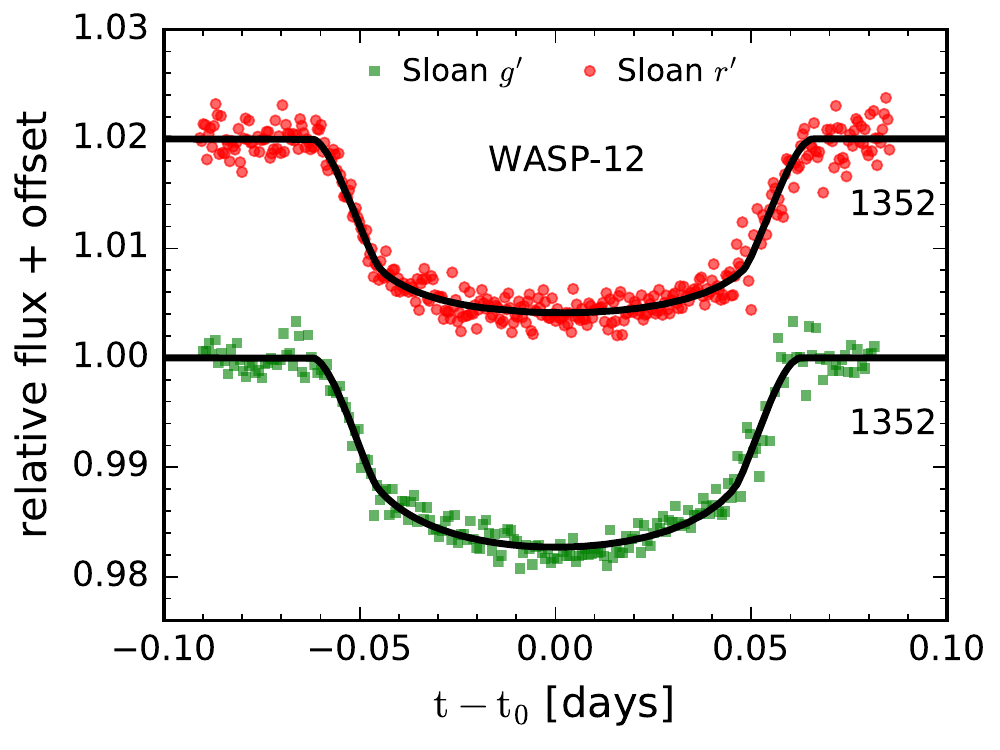}
	\caption{Multicolor light curves of WASP-12 b from epoch 1352. }
	\label{fig:muscat1352}
\end{figure}

\begin{figure}[h]
	\centering
	\includegraphics[width=0.95\linewidth]{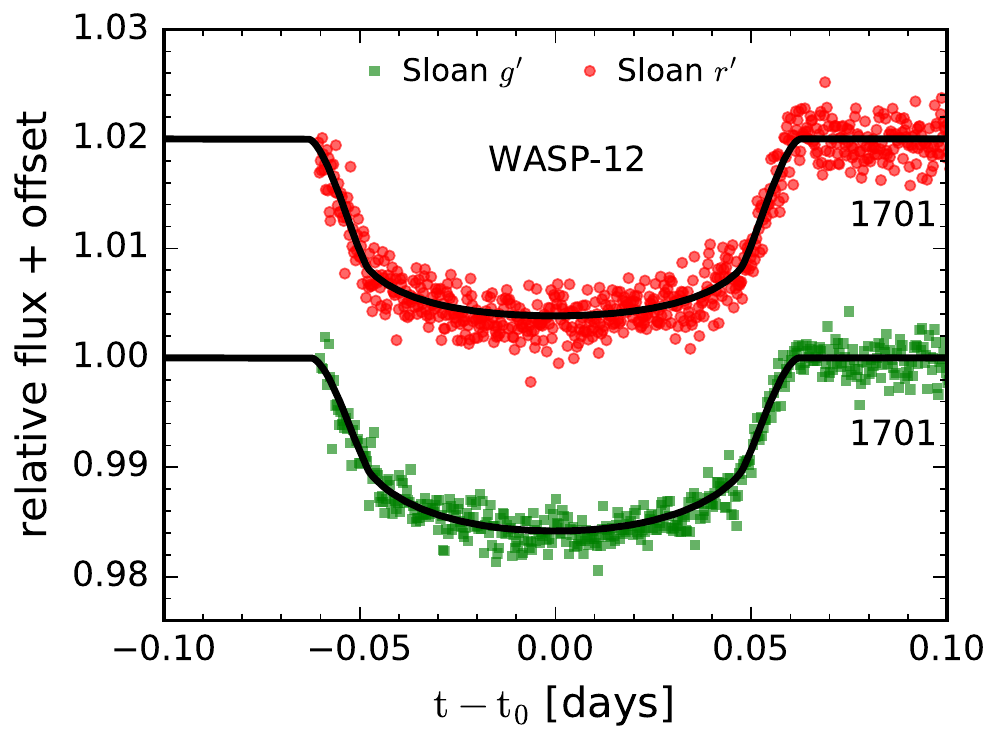}
	\caption{Multicolor light curves of WASP-12 b from epoch 1701.}
	\label{fig:muscat1701}
\end{figure}

\begin{figure}[h]
	\centering
	\includegraphics[width=0.95\linewidth]{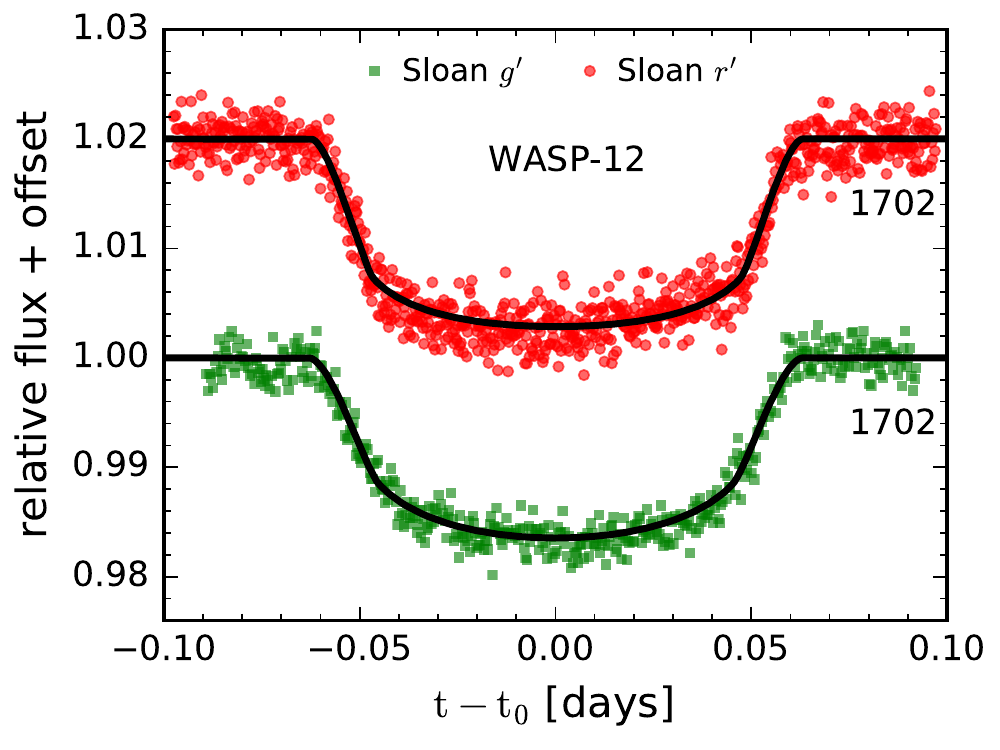}
	\caption{Multicolor light curves of WASP-12 b from epoch 1702.}
	\label{fig:muscat1702}
\end{figure}

\begin{figure}
	\centering
	\includegraphics[width=1.0\linewidth]{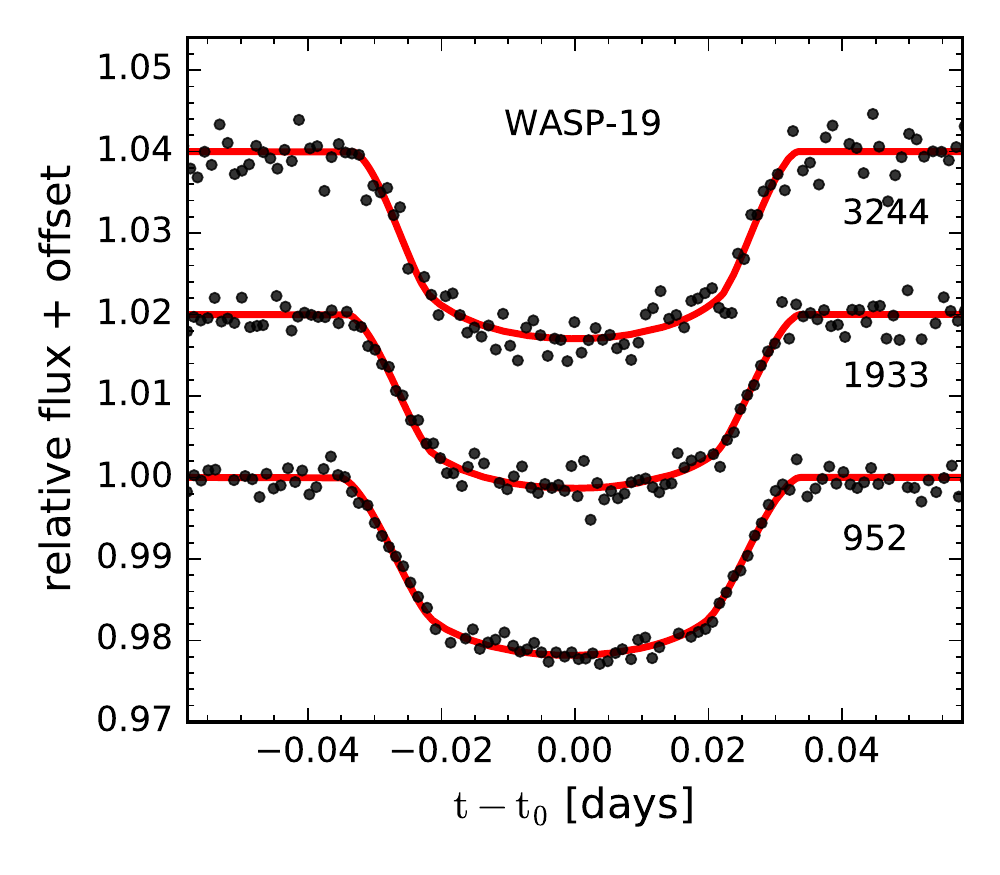}
	\caption{New transit light curves for WASP-19 b. Epoch numbers are printed to the right of each curve. Vertical offsets have been applied to separate the light curves.}
	\label{fig:wasp19lights}
\end{figure}

\begin{figure}
	\centering
	\includegraphics[width=1.0\linewidth]{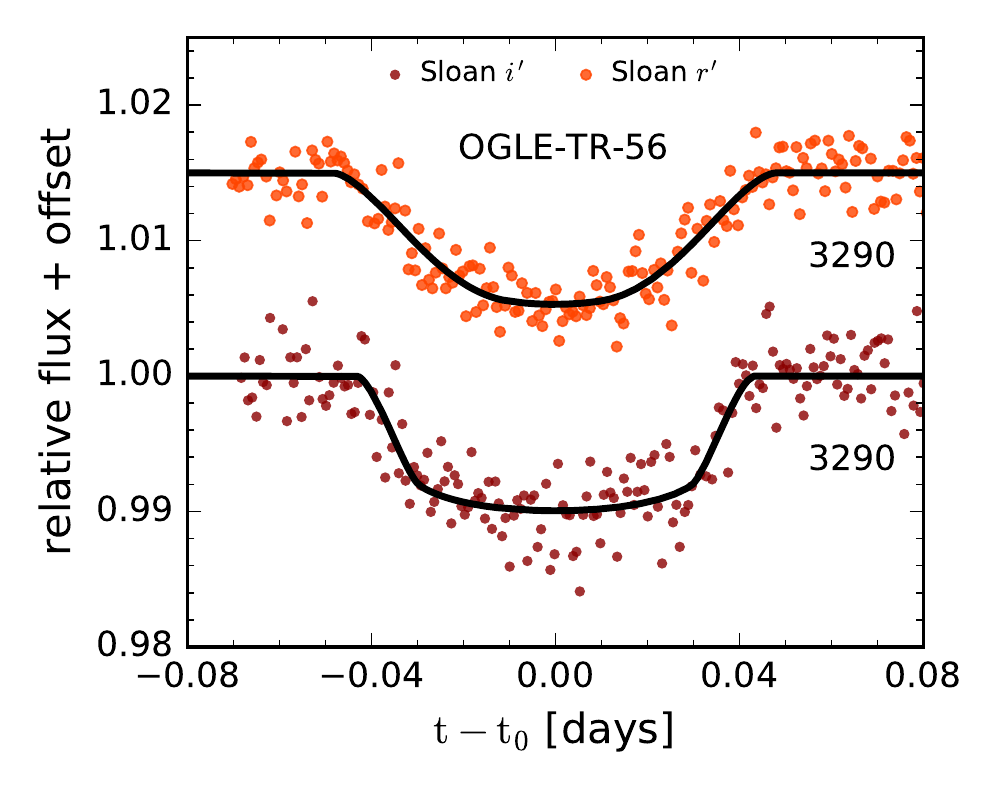}
	\caption{New transit light curves for OGLE-TR-56 b. Epoch numbers are printed to the right of each curve. Vertical offsets have been applied to separate the light curves.}
	\label{fig:ogletr56_light}
\end{figure}

\begin{figure}
	\centering
	\includegraphics[width=1.0\linewidth]{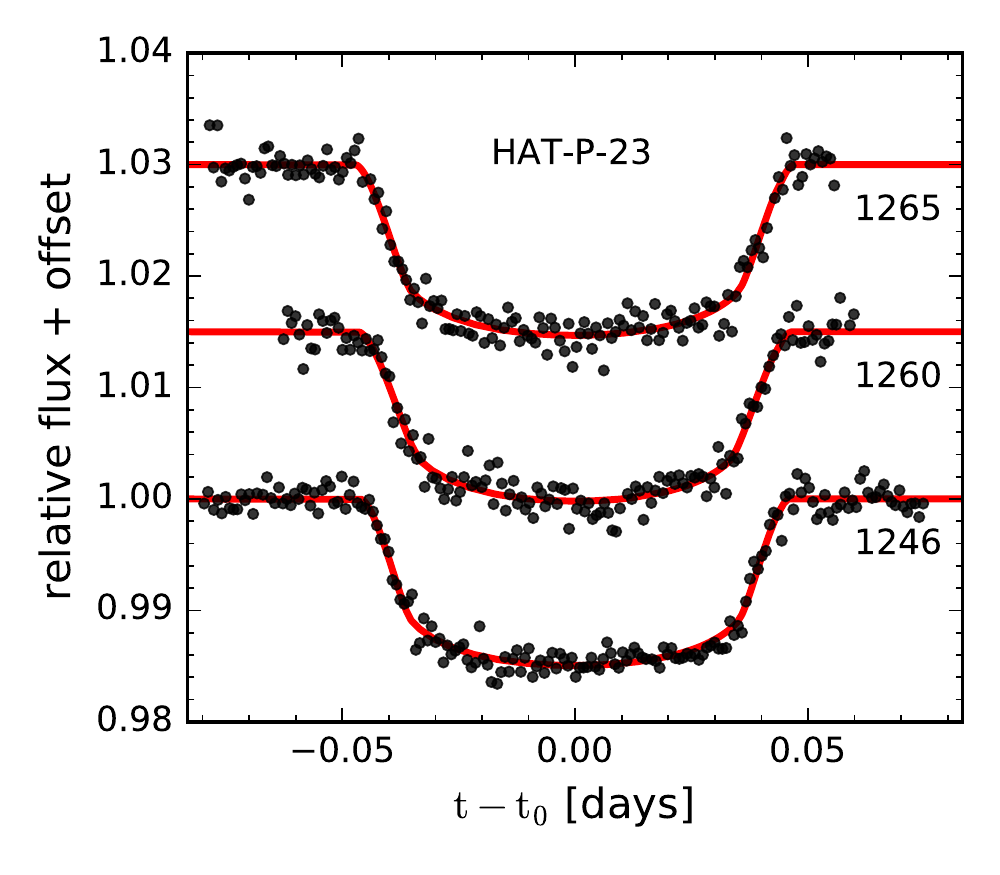}
	\caption{New transit light curves for HAT-P-23 b. Epoch numbers are printed to the right of each curve. Vertical offsets have been applied to separate the light curves.}
	\label{fig:hatp23lights}
\end{figure}

\begin{figure}
	\centering
	\includegraphics[width=1.0\linewidth]{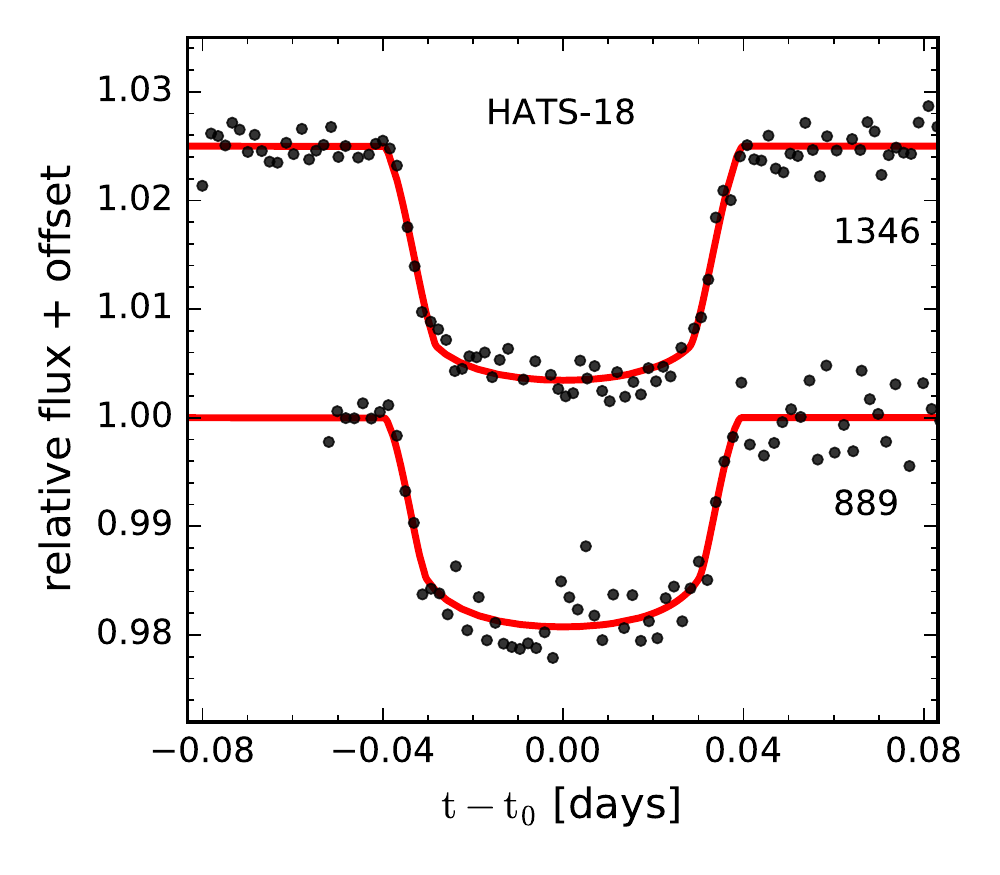}
	\caption{New transit light curves for HATS-18 b. Epoch numbers are printed to the right of each curve. Vertical offsets have been applied to separate the light curves.}
	\label{fig:hats18lights}
\end{figure}

\begin{figure}
	\centering
	\includegraphics[width=1.0\linewidth]{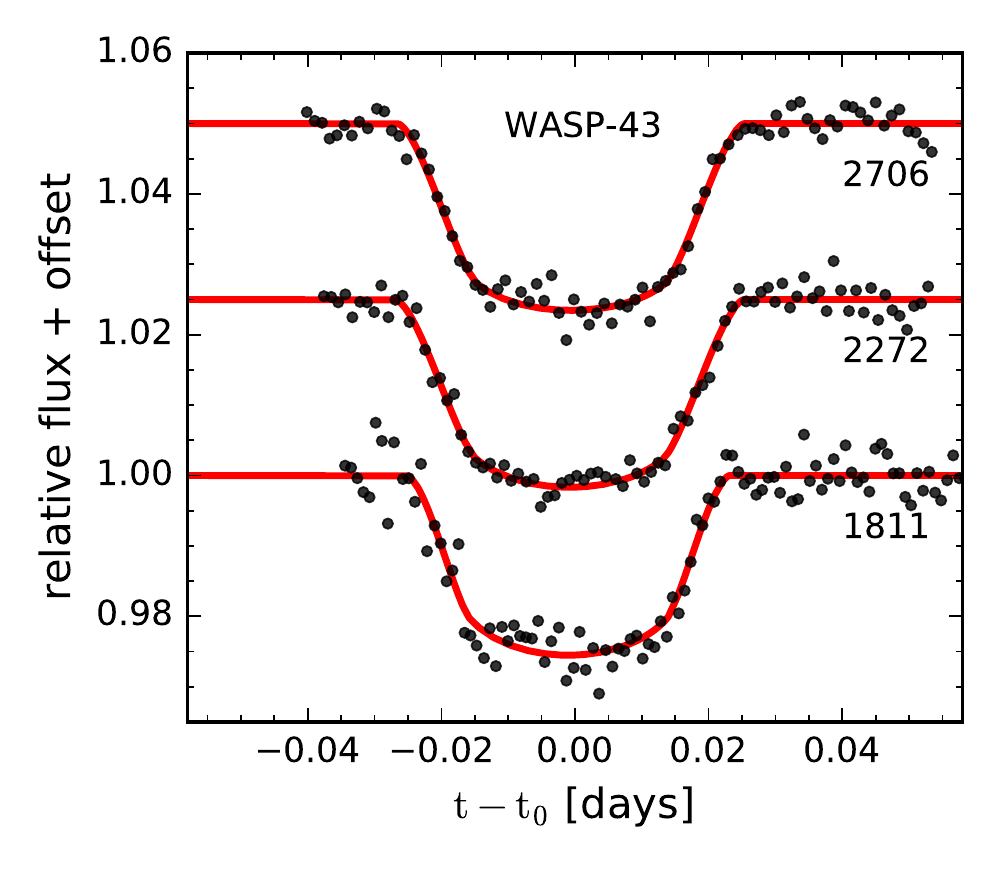}
	\caption{New transit light curves for WASP-43 b. Epoch numbers are printed to the right of each curve. Vertical offsets have been applied to separate the light curves.}
	\label{fig:wasp43lights}
\end{figure}

\begin{figure}
	\centering
	\includegraphics[width=1.0\linewidth]{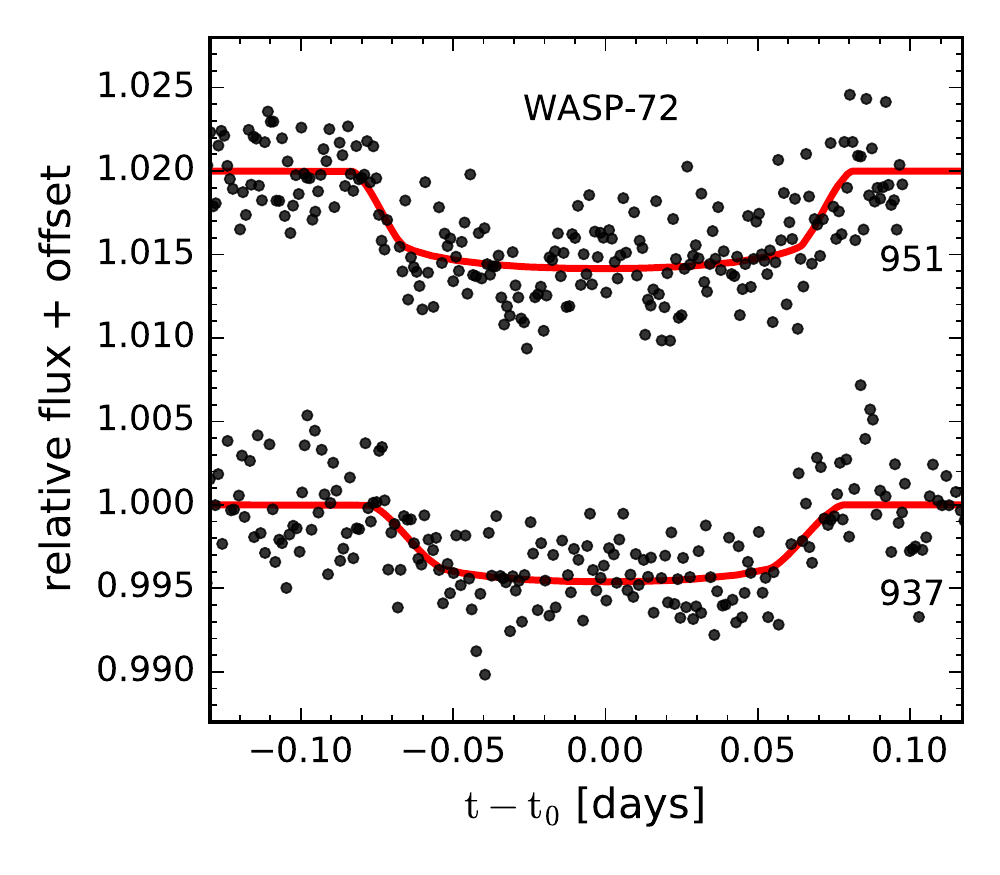}
	\caption{New transit light curves for WASP-72 b. Epoch numbers are printed to the right of each curve. Vertical offsets have been applied to separate the light curves.}
	\label{fig:wasp72lights}
\end{figure}

\begin{figure}
	\centering
	\includegraphics[width=1.0\linewidth]{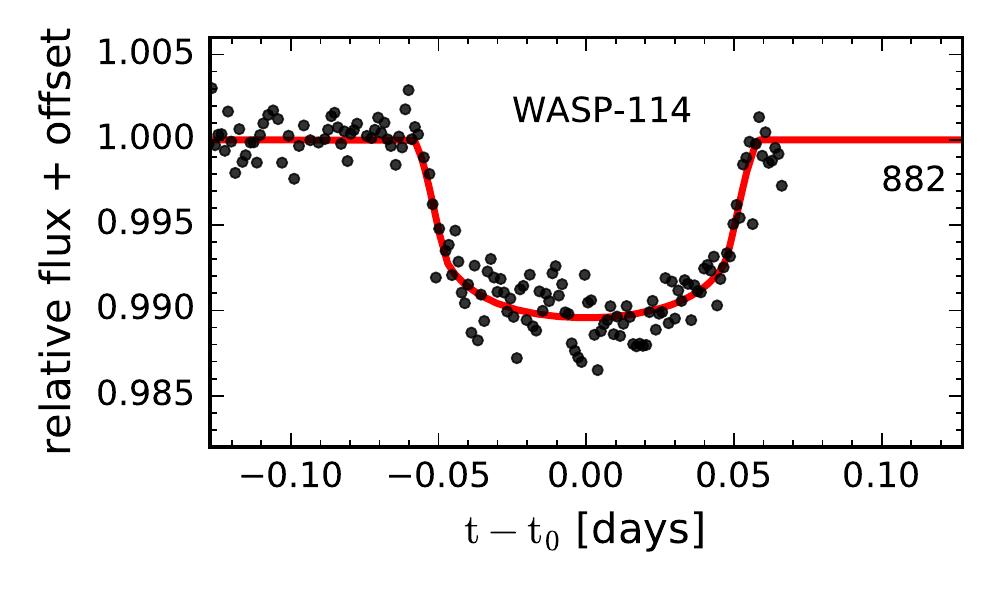}
	\caption{New transit light curve for WASP-114 b. Epoch number is printed to the right of the curve.}
	\label{fig:wasp114lights}
\end{figure}

After each time series was normalized to have unit flux outside of the
transit, we fitted a \citet{MandelAgol2002} model to the data from
each transit. The parameters of the transit model were the mid-transit
time, the planet-to-star radius ratio ($R_{p}$/$R_{\star}$), the
scaled stellar radius ($R_{\star}/a$), and the impact parameter [$b =
a \cos i/R_{\star}$]. For given values of $R_{\star}/a$ and $b$, the
transit timescale is proportional to the orbital period [see, e.g.,
Equation (19) of \citet{Winn2010}]. To set this timescale, we held the
period fixed at its most recent measurement for each target, although
the individual transits were fitted separately with no requirement for
periodicity. To correct for differential extinction, we allowed the
apparent magnitude to be a linear function of airmass, giving two
additional parameters. The limb darkening law was assumed to be
quadratic, with coefficients held fixed at the values tabulated by
\citet{ClaretBloemen2011} for each target star with its spectroscopic
properties adopted from the corresponding discovery
paper. To interpolate the tables, we used the online tool built
by \citet{Eastman+2013}.\footnote{\url{http://astroutils.astronomy.ohio-state.edu/exofast/limbdark.shtml}}
To determine the credible intervals for the parameters, we
used \lstinline{emcee}, an affine invariant Markov Chain Monte Carlo (MCMC) ensemble sampler code
written by \citet{ForemanMackey+2013}. The transition
distribution was proportional to $\exp(-\chi^2/2)$ with
\begin{equation}
\chi^2 = \sum_{i=1}^{N}\left(\frac{f_{{\rm obs},i}-f_{{\rm calc},i}}{\sigma_{i}}\right)^2
\end{equation}
where $f_{{\rm obs},i}$ is the observed flux at time $t_{i}$ and
$f_{{\rm calc},i}$ is the calculated flux based on the model
parameters.  The uncertainties $\sigma_{i}$ were set equal to the
standard deviation of the out-of-transit data. In some transits, the
pre-ingress scatter was noticeably different than the post-egress
scatter; for those observations, we assigned $\sigma_{i}$ by linear
interpolation between the pre-ingress and post-egress values. The
resulting uncertainty estimates appear to be realistic, in the sense
that the measured transit times within a given season generally agree
to within 1-$\sigma$ with a constant period. Hence, no further
allowance for time-correlated noise was made.
Figures \ref{fig:wasp18lights} through \ref{fig:wasp114lights} show
the new light curves and the best-fitting models.
Table~\ref{tbl:times} presents the new transit times for all observed
targets. 

\begin{figure}
	\centering
	\includegraphics[width=1.0\linewidth]{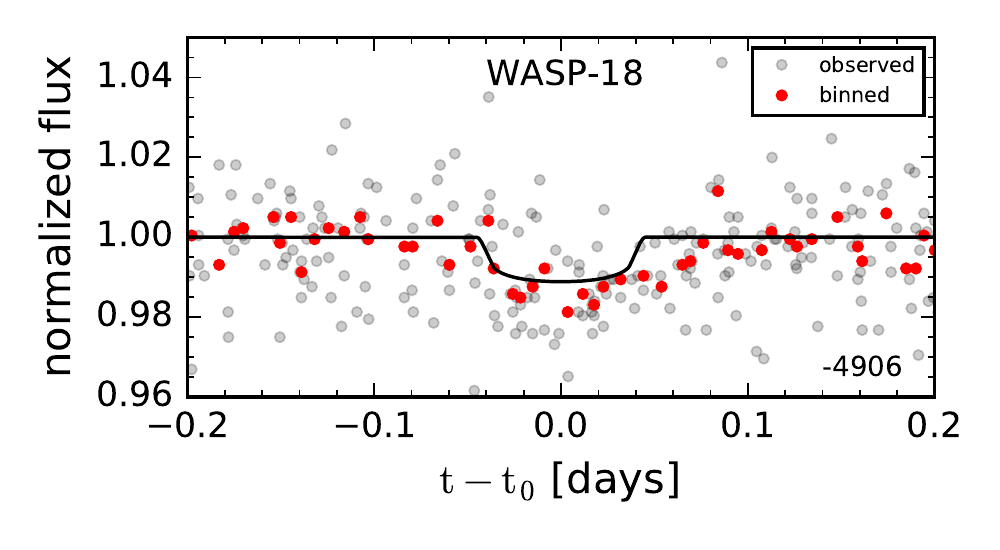}
	\caption{Pre-discovery transit light curve for WASP-18 b. Epoch number is printed to the bottom right of the frame.}
	\label{fig:wasp18lightsasas}
\end{figure}

In addition, to extend the time baseline as far as possible, we looked
for any pre-discovery transit detections in the database of the All
Sky Automated Survey [ASAS; \citet{Pojmanski1997}].  We considered
only the data having a quality grade of A or B (good to average
quality).  We converted the time stamps from HJD\textsubscript{UTC} to
BJD\textsubscript{TDB} using the online applet of
\citet{Eastman2010}. Different datasets were normalized to the same
median magnitude and a ``best aperture'' column was produced that gave
the lowest scatter. Outliers were removed by 5-$\sigma$-clipping. To
retrieve the transit signal, the time-series was phase-folded using
the discovery epoch and orbital period. The only case for which we
detected a significant transit dip was WASP-18, shown in
Figure~\ref{fig:wasp18lightsasas}. The transit was fitted with a model 
following an identical procedure used for other planets. 

\section{Timing Analysis}
\label{sec:timing}

For each target, we considered all the transit times from this work as
well as all the available times reported in the literature for which
(i) the midpoint was allowed to be a completely free parameter in the
fit to the light curve, (ii) the time system was documented clearly,
and (iii) the light curve included both the ingress and the egress of
the transit.  For KELT-16b, WASP-103b and HAT-P-23b, in particular, we
built on the study of \citet{Maciejewski+2018}.  Apart from presenting
new transits for these systems, those authors refitted the transit
data drawn from the literature for the sake of homogeneity. We note
that their reanalyzed times differ from the originally reported times
by several minutes in some cases (see below for more details). Table~\ref{tbl:all_times}
presents the transit times for all targets.

We fitted two models to the timing data.
The first model assumes a constant
orbital period:
\begin{equation}
t_{\rm tra}(E) = t_{0} + PE,
\end{equation}
where $E$ is the epoch number and $P$ is the period. The second model
assumes a constant period derivative:
\begin{equation}
t_{\rm tra}(E) = t_{0} + PE + \frac{1}{2} P \frac{dP}{dt} E^{2}.
\end{equation}
We determined the best-fitting parameters and their uncertainties
using a Markov Chain Monte Carlo (MCMC) method.  Below we summarize
the results for each target.  In all of the figures showing the timing
residuals, with the exception of WASP-114 and HATS-18, we have also plotted a band
depicting the 1-$\sigma$ range of uncertainty in the orbital decay
model. This region encapsulates 68\% of the orbital decay solutions
from the Markov chains.  For WASP-114 and HATS-18, currently only two and three transit times 
are available respectively, so the red band in Figure \ref{fig:wasp114oc} and \ref{fig:hats18oc}
represents the 1-$\sigma$ range of uncertainty in the constant-period
model.

\begin{table*}
\centering
\begin{tabular}{cccccccc}
	\hline 
	\hline
	Target &Date-Obs &Telescope&Filter &Exp  &Epoch& T\textsubscript{mid}& Unc  \\ 
	&  (UTC)  &&&(sec) &  &BJD\textsubscript{TDB} (days) & (days)  \\ 
	\hline 
	
	WASP-18 b &2005-Feb-01 &ASAS& $V$ & 180  &-4906& 2453403.36480 &0.00614  \\
	WASP-18 b &2016-Oct-28 &TRAPPIST-South& $z^\prime$ & 7  &-353& 2457689.79147 &0.00075  \\
	
	\hline
	KELT-16 b &2017-June-10 &FLWO& $r^\prime$ & 30  &5& 2457914.88456 &0.00051  \\ 
	
	KELT-16 b & 2017-June-11 &FLWO& $r^\prime$ &30  &6& 2457915.85370 &0.00062 \\ 
	\hline
		 
	WASP-103 b& 2016-May-04  &FLWO& $r^\prime$  &30 & 731 & 2457512.87006 &0.00031  \\ 
			
	WASP-103 b&  2016-May-30 &FLWO& $r^\prime$ &30 & 759 &2457538.78548 &0.00030 \\ 
			
	WASP-103 b&  2016-Jun-12 &FLWO& $r^\prime$ &30 & 773 &2457551.74314 &0.00038 \\ 
			
	WASP-103 b&  2017-Apr-17 &FLWO& $r^\prime$ &30 & 1107 & 2457860.87503 &0.00029 \\ 
	\hline
	
	WASP-12 b&  2017-Jan-27 &MuSCAT& $g^\prime$ & 30 & 1352 &2457781.05418 &0.00043 \\
	
	WASP-12 b&  2017-Jan-27 &MuSCAT& $r^\prime$ & 15 & 1352 &2457781.05566 &0.00036\\
	
	WASP-12 b&  2018-Feb-12 &MuSCAT& $g^\prime$ & 30 & 1701 &2458161.95991 &0.00035\\
	
	WASP-12 b&  2018-Feb-12 &MuSCAT& $r^\prime$ & 15 & 1701 &2458161.95964 &0.00026\\
	
	WASP-12 b&  2018-Feb-13 &MuSCAT& $g^\prime$ & 30 & 1702 &2458163.05089 &0.00034 \\
	
	WASP-12 b&  2018-Feb-13 &MuSCAT& $r^\prime$ & 15 & 1702 &2458163.05125 &0.00021 \\
	
	\hline
		
	WASP-19 b&  2014-Apr-25  &TRAPPIST-South& $I+z^{\prime}$ & 12  &952 & 2456772.67789 &0.00052  \\ 
			
	WASP-19 b&  2016-Jun-07 & TRAPPIST-South& $I+z^{\prime}$ &12 & 1933 & 2457546.53025 &0.00038 \\ 
			
	WASP-19 b&  2019-Apr-07 &TRAPPIST-South& $I+z^{\prime}$ & 12 &3244 & 2458580.69724 &0.00040 \\ 
	\hline

	HAT-P-23 b&  2016-Sep-09  &FLWO& $r^\prime$ & 20  & 1246 & 2457640.69114 &0.00046  \\ 
			
	HAT-P-23 b&  2016-Sep-26 & FLWO& $r^\prime$ &20 & 1260& 2457657.67134 &0.00062 \\ 
			
	HAT-P-23 b&  2016-Oct-02& FLWO& $r^\prime$ &20 & 1265& 2457663.73660 &0.00044 \\ 
			
	\hline

	WASP-72 b&  2016-Sep-29  &TRAPPIST-South& $I+z^{\prime}$ & 10  & 937 &2457660.73274 &0.00302  \\ 
	WASP-72 b&  2016-Oct-30  & TRAPPIST-South& $I+z^{\prime}$ &10  & 951 &2457691.77307 &0.00250  \\ 
	\hline 
	
	WASP-114 b&  2017-Oct-07  & FLWO& $r^{\prime}$ &30  & 882 &2458033.75834 &0.00068  \\ 
	\hline 
	
	OGLE-TR-56 b &    2017-Jun-19& Magellan Clay & $r^\prime$ &20  & 3290 &2457923.79020 &0.00091  \\ 
	
	OGLE-TR-56 b &    2017-Jun-19& Magellan Clay & $i^\prime$ &20 & 3290 &2457923.78858 &0.00108  \\

	\hline 
	
	WASP-43 b &    2017-Mar-03& TRAPPIST-South & $z^\prime$ &12  & 1811 &2457815.54442 &0.00038  \\ 
	
	WASP-43 b &    2018-Mar-13& TRAPPIST-North & $I+z^\prime$ &10 & 2272 &2458190.55560 &0.00026  \\ 
	
	WASP-43 b &    2019-Mar-01& TRAPPIST-North & $z^\prime$ &12 & 2706 &2458543.60361  &0.00027  \\ 
	
	\hline
	
	HATS-18 b &    2017-Mar-22& TRAPPIST-South & $I+z^\prime$ &30 & 889 & 2457834.74845 &0.00040  \\ 
	
	HATS-18 b &    2018-Apr-09& TRAPPIST-South & $z^\prime$ &25 & 1346 & 2458217.64370 &0.00029  \\ 
	
	\hline 
\end{tabular} 
\caption{Observation journal and measured mid-transit times.} 
\label{tbl:times}
\end{table*}

\begin{table*}
\centering
\begin{tabular}{ccccc}
	\hline 
	\hline
	Target& Epoch& T\textsubscript{mid} (BJD\textsubscript{TDB}) & Error(days) & Reference \\
	\hline
	HAT-P-23&-1163&2454718.84863&0.000466&Bakos et al. 2011\\
	&-1159&2454723.69918&0.000388&''\\
	&-1117&2454774.6412&0.000555&''\\
	&-909&2455026.92065&0.000436&''\\
	&-341&2455715.84171&0.001013&Ram\'{o}n-Fox \& Sada 2013\\
	&-290&2455777.69757&0.001314&''\\
	&-280&2455789.82601&0.001098&''\\

	\hline	

\end{tabular} 
\caption{Transit times of all targets. This table represents the form of timing data; the complete table can be accessed from the electronic version
	of this work.} 
\label{tbl:all_times}
\end{table*}

\begin{enumerate}

\item \textbf{WASP-18\,b} is a ``super-Jupiter'' with a mass of
  11.4~$M_{\rm Jup}$ in a 0.94-day orbit around a relatively hot
  F-type star \citep{Hellier+2009,Stassun+2017}.  The planet's
  mass is large enough that the system is not formally unstable to
  tidal decay, although the period may be shrinking as the system
  approaches synchronization.  \citet{Wilkins+2017} found no signs of
  orbital decay in the system over a decade.
  \citet{McDonaldKerins2018} found 1.3-$\sigma$ (i.e., weak)
  evidence for orbital decay when considering a pre-discovery transit
  observation by Hipparcos.  More recently,
  \cite{Shporer+2019} reported on observations with the {\it
    Transiting Exoplanet Survey Satellite} \citep{Ricker+2015}, adding
  high-precision transit times to the existing ephemeris.  
  
  We compiled and analyzed the available transit timing data, shown in
  Figure~\ref{fig:wasp18oc}. The best-fitting orbital decay model
  gives $\frac{dP}{dt} = -(1.2 \pm 1.3) \times 10^{-10}$. The evidence
  for orbital decay found by \cite{McDonaldKerins2018} has been
  further weakened with the inclusion of the {\it TESS} data.  Due to
  a lack of conclusive evidence for orbital decay, we can set a
  95\%-confidence lower limit on $Q'_{\star}$ of $(1.7\pm 0.4) \times
  10^6$.  Here and elsewhere, this limit was obtained by calculating
  the 95\%-confidence limit on $dP/dt$ and then applying Equation~(1)
  to convert it into a limit on $Q'_{\star}$.  The uncertainty of 0.4
  is based on the propagation of the uncertainties in $M_{\rm
    p}/M_\star$ and $a/R_\star$ from Table 1.

  Our result for WASP-18 is consistent with the 2-$\sigma$
  limit $Q'_{\star} > 10^{6}$ derived by \cite{Wilkins+2017}.
  The refined transit ephemeris is $t_{\rm tra}(E) = 2458022.12523(02) ~\rm{
    BJD_{TDB}}$ + $E \times 0.941452425(22)~ \rm{ days}$.

\begin{figure*}
	\centering
	\includegraphics[width=1.0\linewidth]{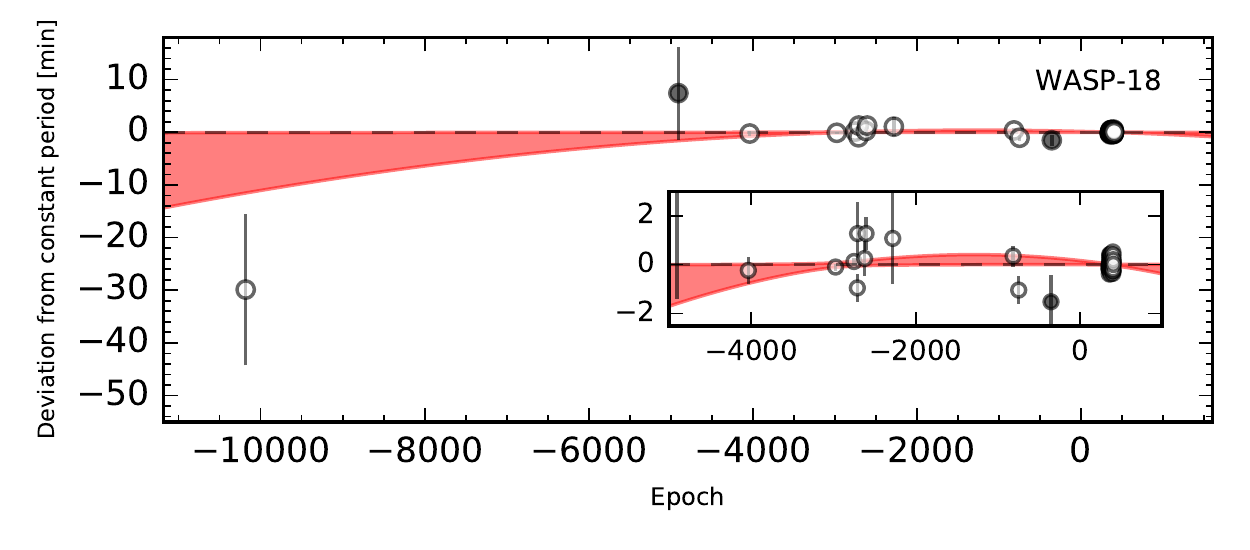}
	\caption{Timing residuals for WASP-18\,b. White
          points represent transit times collected from the
          literature: \citet{Hellier+2009}, \citet{Maxted+2013},
          \citet{Wilkins+2017}, \citet{McDonaldKerins2018} and \citet{Bouma+2019}. The black point shows the pre-discovery
          transit from ASAS. The inset, with same axes units as the
          main frame, shows a zoomed in version of timing residuals
          between epochs $-5000$ and 1000. The red band represents
          the 1-$\sigma$ uncertainty in the orbital decay model.}
	\label{fig:wasp18oc}
\end{figure*}

\item \textbf{KELT-16\,b} is a 2.8~$M_{\rm Jup}$ planet in a
  0.97-day orbit around an F star \citep{Oberst+2017}. In their
  follow-up study, \citet{Maciejewski+2018} refitted the transits
  presented by \citet{Oberst+2017} to redetermine transit times. Their
  times differ from those reported originally by about 30 seconds. In
  this work, we present two new transit times, which put together with
  times reported by \citet{Maciejewski+2018} are shown in Figure
  \ref{fig:kelt16oc}

\begin{figure*}
	\centering
	\includegraphics[width=1.0\linewidth]{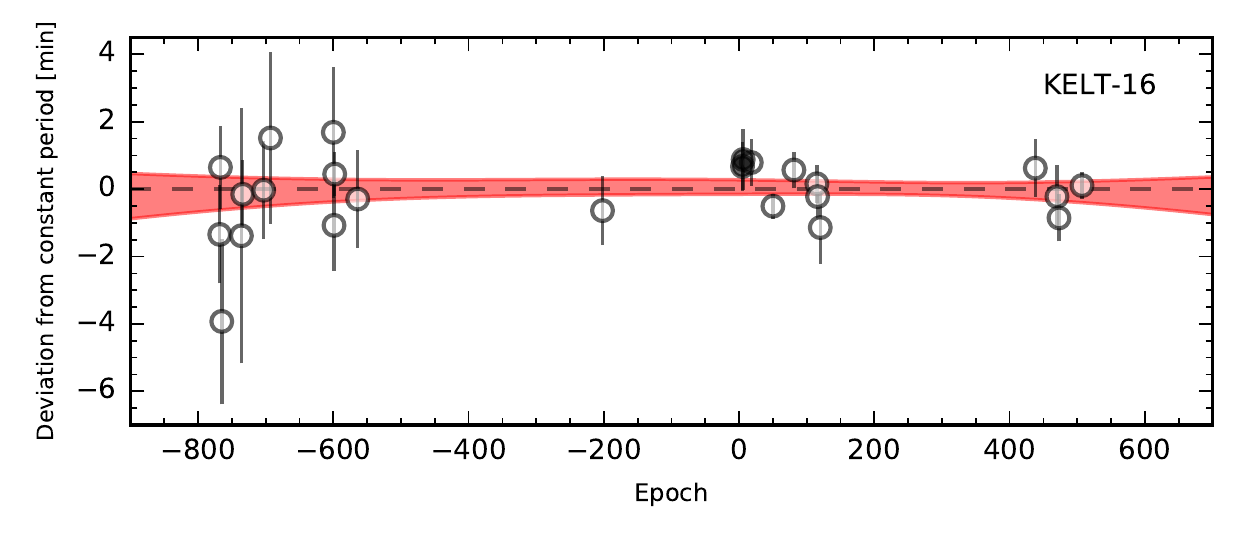}
	\caption{Timing residuals for KELT-16b. White points represent transit times originally reported by \citet{Oberst+2017} and reanalyzed by \citet{Maciejewski+2018}. Black points represent new data. The red band represents 1-$\sigma$ uncertainty on the orbital decay model.}
	\label{fig:kelt16oc}
\end{figure*}

The best-fit orbital decay model results in $\frac{dP}{dt} = -(0.6 \pm
1.4) \times 10^{-9}$, consistent with a constant period.  The lack of
a decreasing period allows us to limit the reduced tidal quality factor to
$Q'_{\star} > (0.9\pm 0.2) \times 10^5$ with 95\% confidence.  We
refined the constant-period ephemeris to $t_{\rm tra}(E) =
2457910.03913(11) ~\rm{ BJD_{TDB}}$ + $E \times 0.96899319(30)~ \rm{
  days}$.

\item \textbf{WASP-103\,b} is a 1.5~$M_{\rm Jup}$ planet in a
  0.93-day orbit around a late F star \citep{Gillon+2014}.
  \cite{Maciejewski+2018} presented several new transit times and also
  reanalyzed light curves available in the literature.  We present 4
  new transits of WASP-103b which, along with times from
  \cite{Maciejewski+2018}, are shown in Figure~\ref{fig:wasp103oc}.

\begin{figure*}
	\centering
	\includegraphics[width=1.0\linewidth]{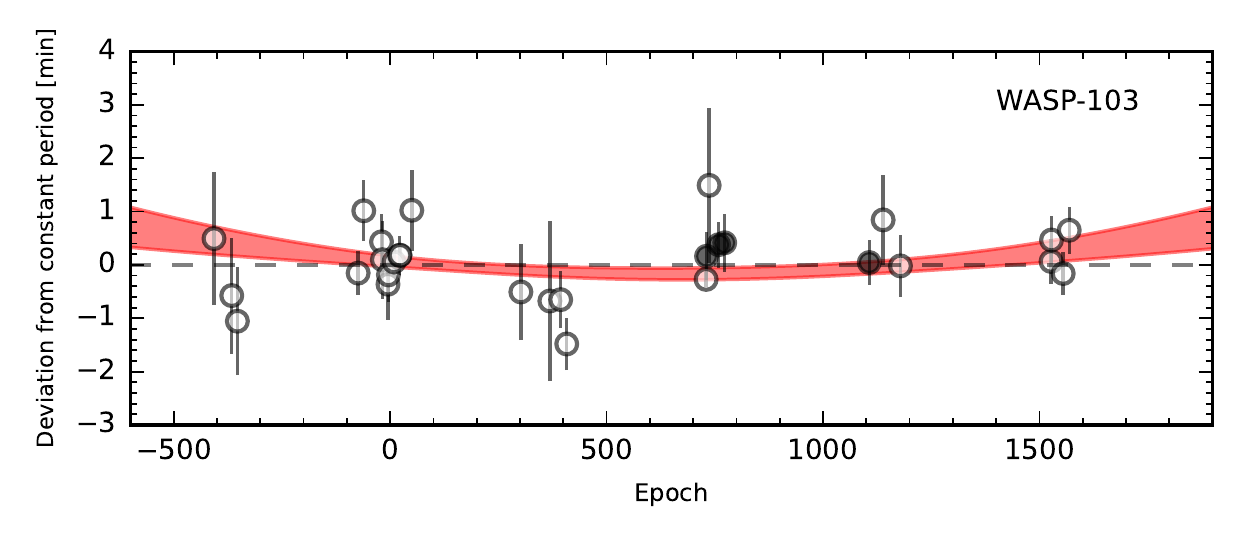}
	\caption{Timing residuals for WASP-103\,b. White points are
          data obtained and reanalyzed by \cite{Maciejewski+2018} and
          originally reported by \cite{Gillon+2014, Southworth+2015,
            Delrez+2018, Turner+2017} and \cite{Lendl+2017}. Black
          points represent new data. The red band represents the
          1-$\sigma$ uncertainty inn the orbital decay model.}
	\label{fig:wasp103oc}
\end{figure*}

The best-fit orbital decay model results in $\frac{dP}{dt} = (8.42 \pm
3.95) \times 10^{-10}$. Note the positive period derivative, which
could be a statistical artifact or the sign of a period change due to
some other reason (such as a third body in the system, or apsidal
precession).  Treating the result as a non-detection of orbital decay,
we can constrain the tidal quality factor to be $Q'_{\star} > (1.1\pm 0.1) \times 10^{5}$ with 95\% confidence. We refined the linear ephemeris
to $t_{\rm tra}(E) = 2456836.29630(07) ~\rm{ BJD_{TDB}}$ + $E \times
0.925545352(94)~\rm{days}$.

\item \textbf{WASP-12\,b} is a 1.5~$M_{\rm Jup}$ planet in 1.1-day
  orbit around a star that appears to be a late F main-sequence star,
  but could also be a subgiant \citep{Hebb+2009,Weinberg2017}.
  \cite{Maciejewski+2016} reported the first indication of a
  systematic quadratic deviation from a constant-period model.
  \cite{Patra+2017} confirmed the aforementioned result and found the
  rate of period change to be $-29 \pm
  3$~ms~yr\textsuperscript{-1}.\footnote{See also
    \url{http://var2.astro.cz/ETD/etd.php?STARNAME=WASP-12&PLANET=b}.}
  The implied tidal quality factor is $Q_{\star} \approx 2 \times 10^5$.
  Although both sets of authors pointed out that the available data might also be
  explained by the apsidal precession of the orbit, \cite{Yee+2020}
  found that the interval between the measured times of occultations
  has also been shrinking, which is evidence in favor of orbital decay.

The sky-projected obliquity of WASP-12\,b has been measured to be
$\lambda = 59^{+15}_{-20}$~deg. The sky-projected rotation velocity
$v\sin i$ is less than 2.2~km~s$^{-1}$, which is unusually low for an F-type star
\citep{Albrecht+2012, Hebb+2009}, and suggests that our line of sight
may be nearly aligned with the stellar rotation axis.  This made us
wonder if the measured transit times are being thrown off from the
true values due to stellar activity.  A slowly-varying spot pattern
near one of the rotation poles could lead to a persistent deformity of
the transit light curve, which would manifest itself as a perturbation
in the fitted transit time. A long-term activity cycle might cause
this perturbation to gradually change amplitude, leading to a spurious
detection of a change in orbital period. We decided to test for this
effect by observing the same transit at different
wavelengths. Starspot activity is generally chromatic. If the apparent
transit time appears to vary with wavelength, that would be a sign of
systematic errors due to stellar activity.

We used the multicolor camera on MuSCAT to observe transits
simultaneously in three different filters: $g^{\prime}$, $r^{\prime}$,
and $z^{\prime}$.  Transits were observed on 2017~Jan~27, 2018~Feb~12,
and 2018~Feb~13, although in practice we used only the $g^{\prime}$
and $r^{\prime}$ data because of the relatively lower quality of the
$z^{\prime}$ data.  All the light curves were fitted
independently. Along with the transit model, we also fitted for linear
functions to decorrelate the apparent magnitude with the $X$ and $Y$
positions of the target star on the CCD, and on the airmass.
Figures~\ref{fig:muscat1352}, \ref{fig:muscat1701} and
\ref{fig:muscat1702} show the light curves and the best-fit transit
models. Figure~\ref{fig:wasp12oc_muscat} shows the positions of these
mid-transit times among all other timing residuals.

\begin{figure*}
	\centering
	\includegraphics[width=1.0\linewidth]{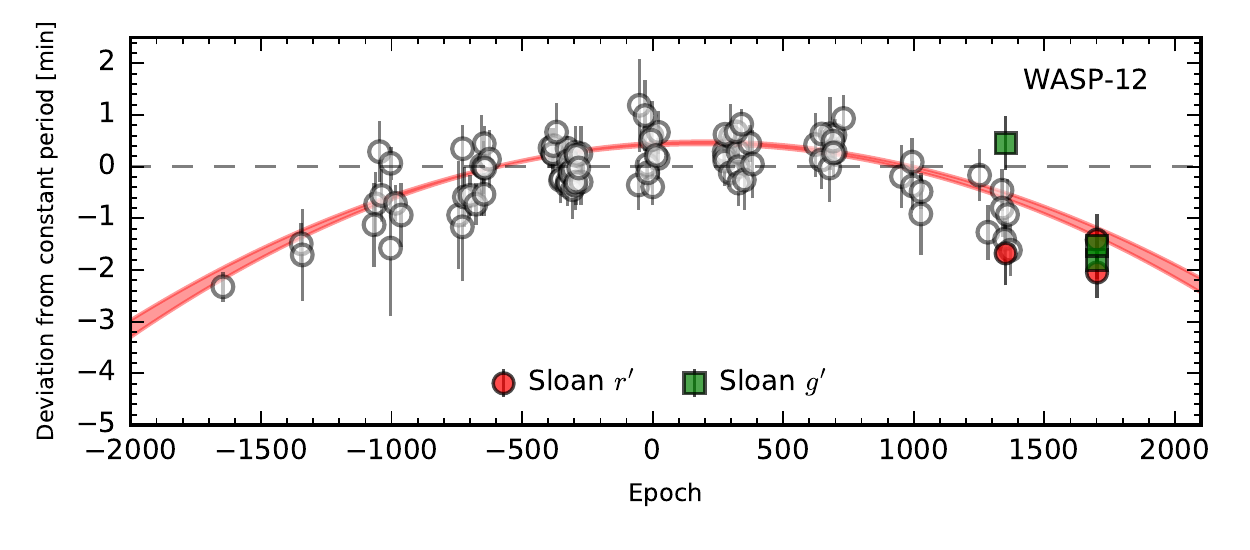}
	\caption{Timing residuals for WASP-12\,b. White points are from \cite{Hebb+2009, Copperwheat+2013, Chan+2011, Collins+2017, Maciejewski+2013,  Sada+2012,  Cowan+2012, Stevenson+2014, Maciejewski+2016, Kreidberg+2015, Patra+2017}. Red and green points represent new
          data obtained with MuSCAT. The red band represents 1-$\sigma$
          uncertainty in the orbital decay model.}
	\label{fig:wasp12oc_muscat}
\end{figure*}

For epoch 1352, the $r^{\prime }$ and $g^{\prime}$ times differ from
each other by 3-$\sigma$. However, we do not ascribe too much
significance to this discrepancy, because on this night there were
very large transparency variations during egress and several data
points had to be excluded as large outliers. The other two transits
were obtained in good weather and the results show good agreement
between the transit times measured in the two bands, as well as
consistency with the previously established trend of a shrinking
period.  Thus, there is no indication that the transit timing
deviations are chromatic, although we do intend to continue multicolor
observations to check further.

\item \textbf{HATS-18\,b} is a 2~$M_{\rm Jup}$ planet in a 0.84-day
  orbit around a G star \citet{Penev+2016}. Since there are only three
  available data points, we were not able to perform a meaningful
  search for period changes, although we were able to reduce the
  uncertainty in the orbital period by a factor of two.  The
  up-to-date ephemeris is $t_{\rm tra}(E) = 2457089.90665(25)
  ~\rm{BJD_{TDB}}$ + $E \times 0.83784309(28)~ \rm{ days}$.

\begin{figure*}
	\centering
	\includegraphics[width=1.0\linewidth]{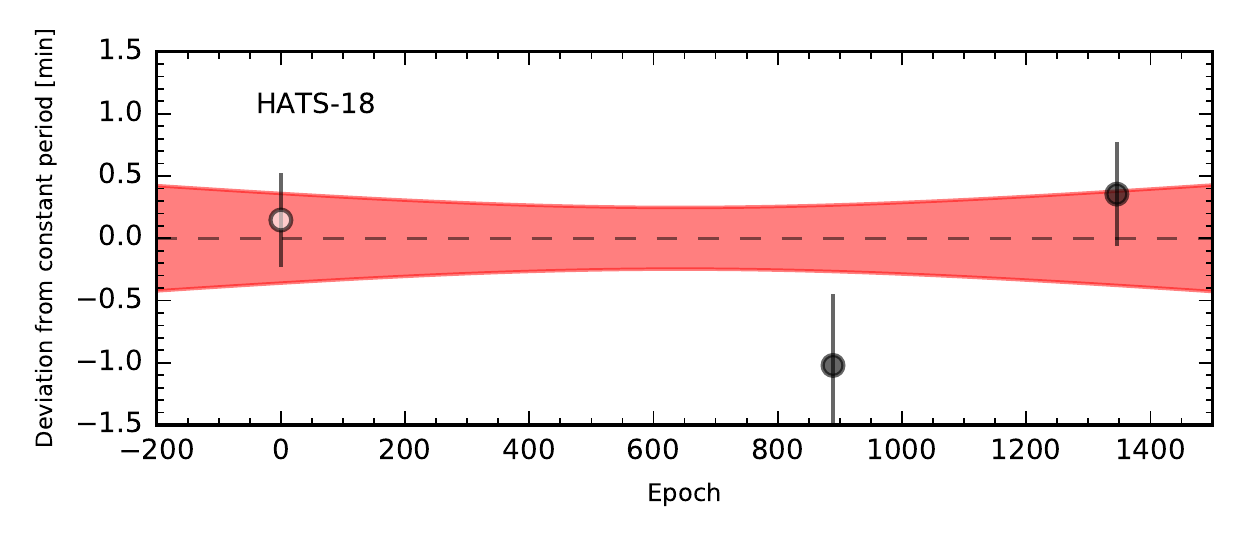}
	\caption{Timing residuals for HATS-18\,b. The white point is from
		\citet{Penev+2016} and the black points represent new
		data. The red band depicts 1$\sigma$ uncertainty on the
		linear ephemeris.}
	\label{fig:hats18oc}
\end{figure*}

\item \textbf{WASP-19\,b} is the shortest-period hot Jupiter yet
  discovered, with a period of 0.78 days \citep{Hebb+2010}. The planet
  has a mass of 1.1~$M_{\rm Jup}$ and orbits a star very similar to
  the Sun in mass.  We have added 3 new transit times to the existing
  collection.  The timing residuals are shown in Figure~\ref{fig:wasp19oc}.

\begin{figure*}
	\centering
	\includegraphics[width=1.0\linewidth]{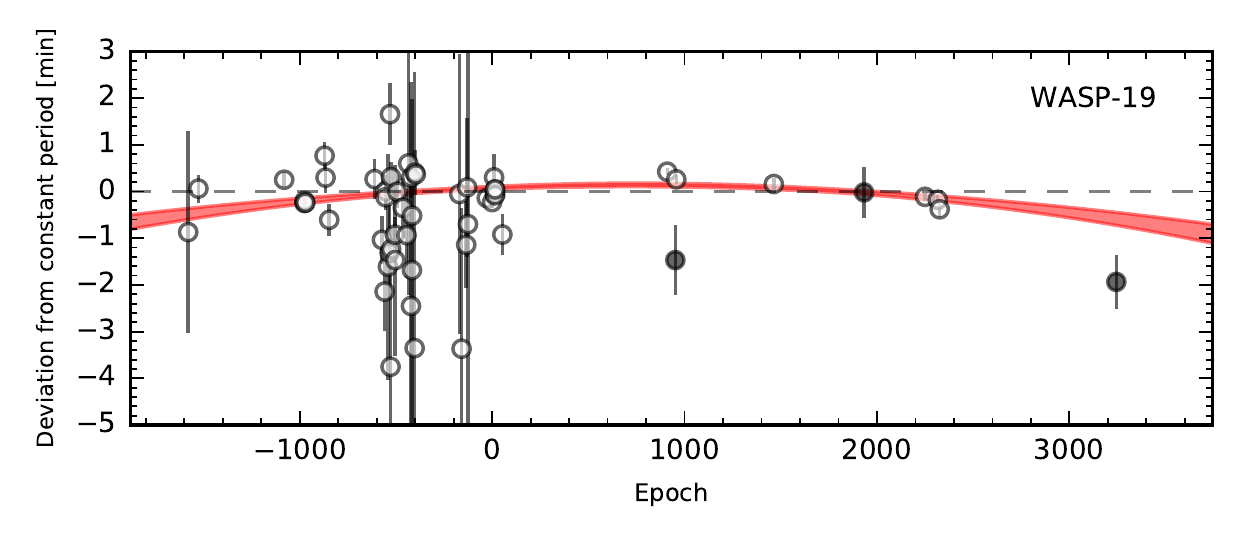}
	\caption{Timing residuals for WASP-19\,b. White points are data
          reported in the literature: \cite{Hebb+2010, Hellier+2011,
            Dragomir+2011, Lendl+2013, TregloanReed+2013,
            Mancini+2013, Bean+2013, Espinoza+2019}. Black points
          represent new data obtained by TRAPPIST. The red band
          represents 1$\sigma$ uncertainty on the orbital decay
          model.}
	\label{fig:wasp19oc}
\end{figure*}

The best-fitting orbital decay model results in $\frac{dP}{dt} =
-(2.06 \pm 0.42) \times 10^{-10}$, equivalent to $-(6.5 \pm
1.3)$~ms~yr$^{-1}$.  If the orbit is indeed decaying, the implied
tidal quality factor is $Q'_{\star} = (5.0\pm 1.5) \times 10^{5}$. For
comparison, WASP-12 has a period derivative of $-~(29 \pm 3)$ ms
yr\textsuperscript{-1} and a reduced tidal quality factor of $(1.6\pm
0.2) \times 10^{5}$.  Formally, the period change of WASP-19\,b is
stastically significant.  However, we are cautious because the data
are relatively scanty in comparison with WASP-12\,b, and there appears
to be a lot of scatter. Two of our new transit times, in particular,
deviate from the best-fitting model by more than 1-$\sigma$.

Both \cite{Mancini+2013} and \cite{Espinoza+2019} attributed some
previously noted timing inconsistencies to starspot activity. In
several cases, an anomaly in the light curve was clearly detected.
For this reason, we believe that further observations are required to
determine if there is a indeed a departure from constant period. In
the meantime, we have refined the linear ephemeris for transits to be
$t_{\rm tra}(E) = 2456021.70390(02) ~\rm{ BJD_{TDB}}$ + $E \times
0.788839300(17)~ \rm{days}$.

\item \textbf{OGLE-TR-56\,b} is the most challenging target in our
  list to observe because of the relatively faint star ($V$ =
  16.6). On the other hand, it is one of the earliest discovered
  transiting exoplanets, which means that there is an unusually long
  time baseline over which data are available. The discovery of this
  planet also inspired some early attempts to model tidal orbital
  decay \citep[see, e.g.,][]{Sasselov2003,CaronePatzold2007}.
  
  \cite{Adams+2011} performed a major transit-timing study based on 21
  light curves. To this, we have added one new transit time in 2017,
  more than 6 years later.  The timing residuals are shown in
  Figure~\ref{fig:ogletr56oc}. We do not find significant evidence for
  orbital decay.  The best-fit orbital decay model results in
  $\frac{dP}{dt} = (3.34 \pm 1.49) \times 10^{-10}$, a weak period
  increase.  Treating this result as a null detection, we conclude
  that $Q'_{\star} > (7.0\pm 2.1) \times 10^{5}$ with 95\%
  confidence. The updated linear ephemeris is $t_{\rm tra}(E) =
  2453936.60063(14) ~\rm{ BJD_{TDB}}$ + $E \times 1.21191123(16)~ \rm{
    days}$.

\begin{figure*}
	\centering
	\includegraphics[width=1.0\linewidth]{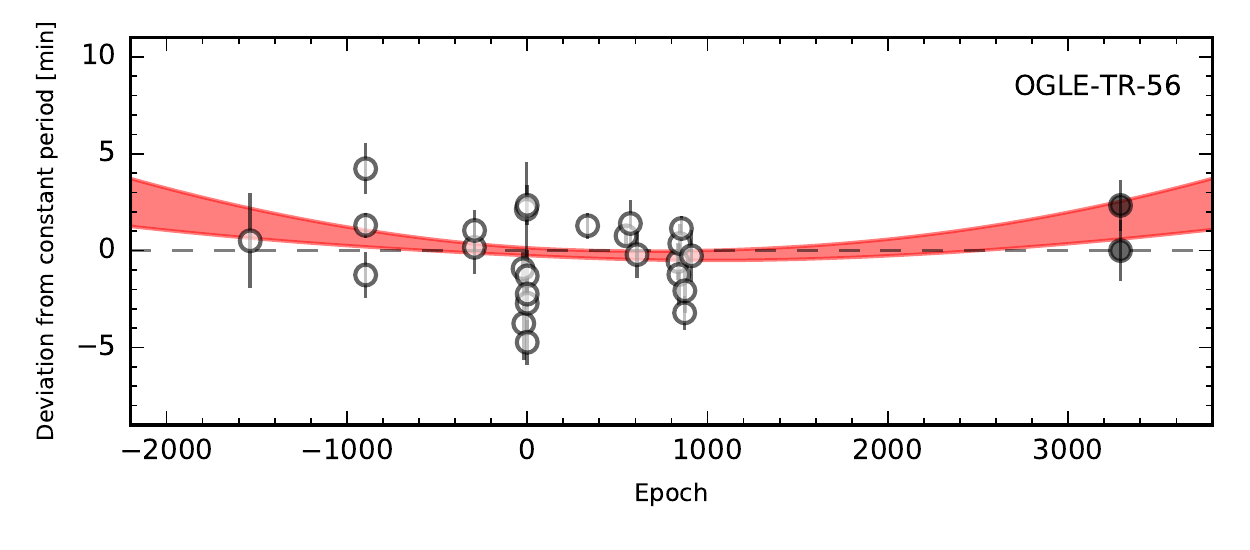}
	\caption{Timing residuals for OGLE-TR-56\,b. White points were
          reported by \cite{Torres+2004, Pont+2007} and
          \cite{Adams+2011}. Black points are the new data. The red
          band represents the 1-$\sigma$ uncertainty in the orbital
          decay model.}
	\label{fig:ogletr56oc}
\end{figure*}

\item \textbf{HAT-P-23\,b} is a 2.1~$M_{\rm Jup}$ planet in a 1.2-day
  orbit around a G dwarf \citep{Bakos+2011}.  \cite{Maciejewski+2018}
  presented several new transit times and renanalyzed some light
  curves drawn from the literature. Their revised transit times differ
  from those reported by \cite{Ciceri+2015} by several minutes, for
  unknown reasons.  We have adopted their reanalyzed transit times,
  figuring that the more homogeneous analysis was preferable.  To
  these data, we have added 3 new transit times. The timing residuals are
  shown in Figure \ref{fig:hatp23oc}.

\begin{figure*}
	\centering
	\includegraphics[width=1.0\linewidth]{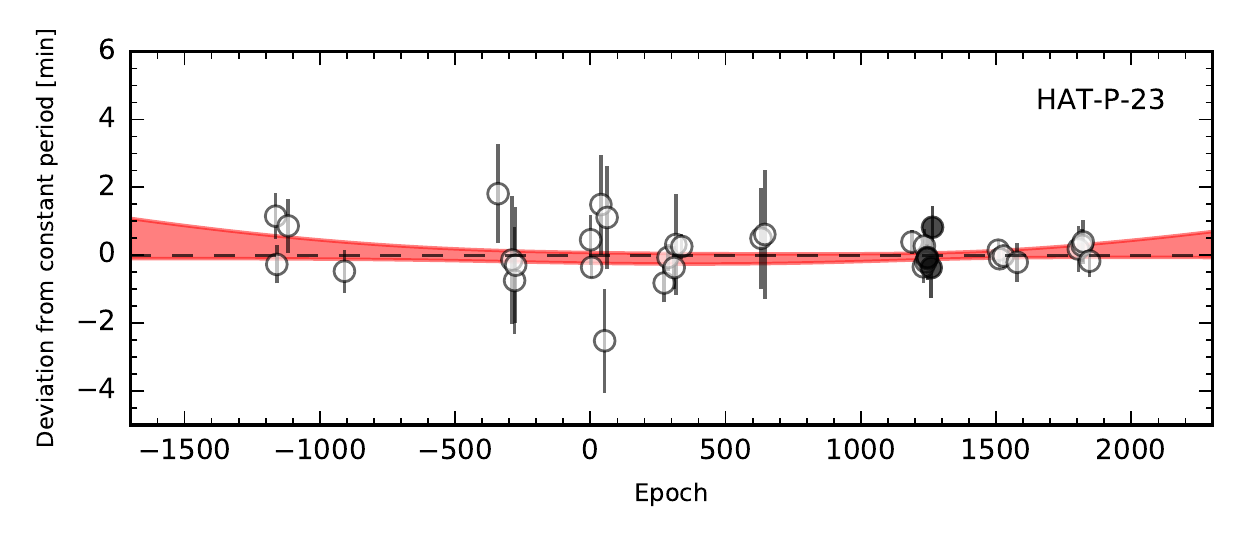}
	\caption{Timing residuals for HAT-P-23\,b. White points are
          data obtained and reanalyzed by \cite{Maciejewski+2018} and
          originally reported by \cite{Bakos+2011, RamonFoxSada2013,
            Ciceri+2015} and \cite{SadaRamonFox2016}. Black points
          represent new data. The red band represents the 1-$\sigma$
          uncertainty in the orbital decay model.}
	\label{fig:hatp23oc}
\end{figure*}

The best-fit orbital decay model results in $\frac{dP}{dt} = (1.47 \pm
1.48) \times 10^{-10}$, consistent with a constant period. The lack of
orbital decay in HAT-P-23b allows us to place a lower limit of
$Q'_{\star} > (6.4\pm 1.9) \times 10^{5}$ with 95\% confidence. We
refined the linear ephemeris to $t_{\rm tra}(E) = 2456129.43472(09)
~\rm{ BJD_{TDB}}$ + $E \times 1.212886411(72)~ \rm{ days}$.

\item \textbf{WASP-72\,b} was discovered by \cite{Gillon+2013}.  It is
  a 1.5~$M_{\rm Jup}$ planet in a 2.2-day orbit around an F star.  We
  have added 2 new data points to the existing collection.  Because
  WASP-72 has a transit depth of only 0.42\%, the timings have
  relatively low precision.

\begin{figure*}
	\centering
	\includegraphics[width=1.0\linewidth]{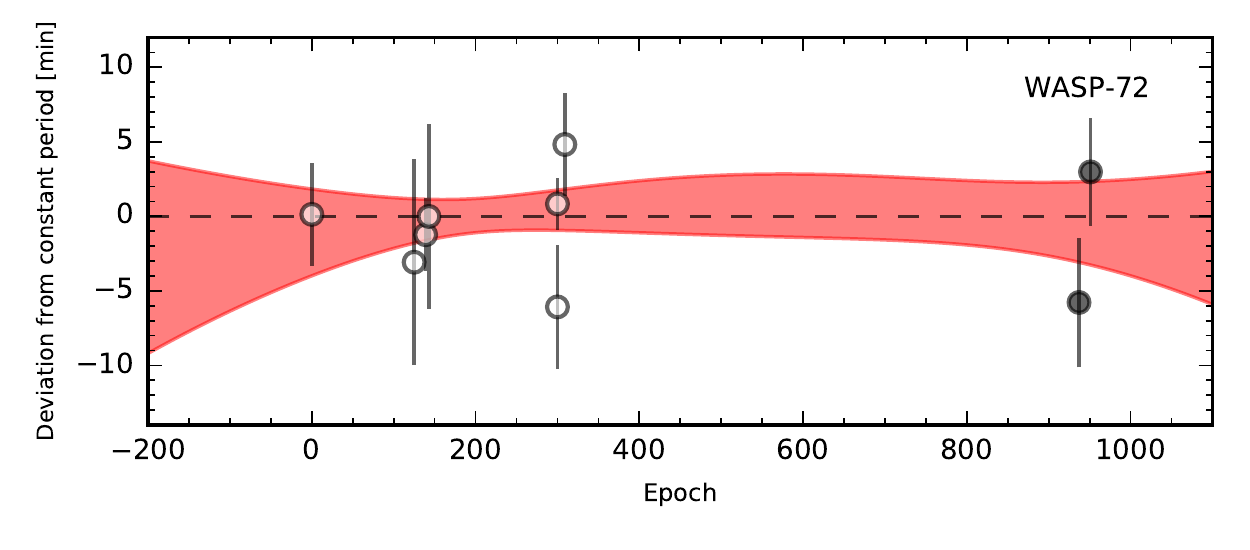}
	\caption{Timing residuals for WASP-72\,b. White points were
          reported by \cite{Gillon+2013}. Black points represent new
          data. The red band represents the 1-$\sigma$ uncertainty in
          the orbital decay model.}
	\label{fig:wasp72oc}
\end{figure*}

The best-fit orbital decay model results in $\frac{dP}{dt} = -(5.7 \pm
9.2) \times 10^{-9}$, consistent with a constant period. WASP-72 has
not been observed for long enough for us to put an interesting limit
on the tidal quality parameter.  The available data lead to the
relatively weak constraint of $Q'_{\star} > (2.1\pm 1.4)\times 10^3$
with 95\% confidence.  We can, however, reduce the uncertainty in the
orbital period by a factor of 3.  The refined linear ephemeris is
$t_{\rm tra}(E) = 2455583.6552(11) ~\rm{BJD_{TDB}}$ + $E \times
2.2167360(27)~ \rm{ days}$.

\item \textbf{WASP-43\,b}, a 2\,$M_{\rm Jup}$ planet in an 0.8-day
  orbit around a K dwarf, was discovered by \cite{Hellier+2011}. Since
  then, it has had an interesting story as far as orbital decay is
  concerned. \cite{Jiang+2016} reported a detection of period
  shrinkage with $\frac{dP}{dt} = -29 \pm 7$~ms~yr$^{-1}$, nominally
  a 4-$\sigma$ detection. However, in the same year,
  \cite{Hoyer+2016} ruled out orbital decay at that rate based on
  additional transit observations.  \cite{Stevenson+2017} added 3
  transits observed with the \textit{Spitzer} Space Telescope and
  found no evidence for orbital decay. We have significantly increased the time baseline
  by adding 3 new transits. In Figure \ref{fig:wasp43oc},
  we compile the transit times after removing some data points based on
  incomplete transits.  We find $\frac{dP}{dt} = (1.9 \pm 0.6)
  \times 10^{-10}$. It appears the period of WASP-43 has increased
  slightly, although there are a lot of outlying data points. Whether this is
  just a statistical fluke will be clearer after more observations in
  the future. Assuming that the period is not changing, we can set a limit 
  of $Q'_\star > (2.1\pm 1.4) \times 10^{5}$ with 95\%
  confidence. The refined linear ephemeris is $t_{\rm tra}(E) =
  2456342.34259(01) ~\rm{ BJD_{TDB}}$ + $E \times 0.813474059(29)~
  \rm{ days}$.

\begin{figure*}
	\centering
	\includegraphics[width=1.0\linewidth]{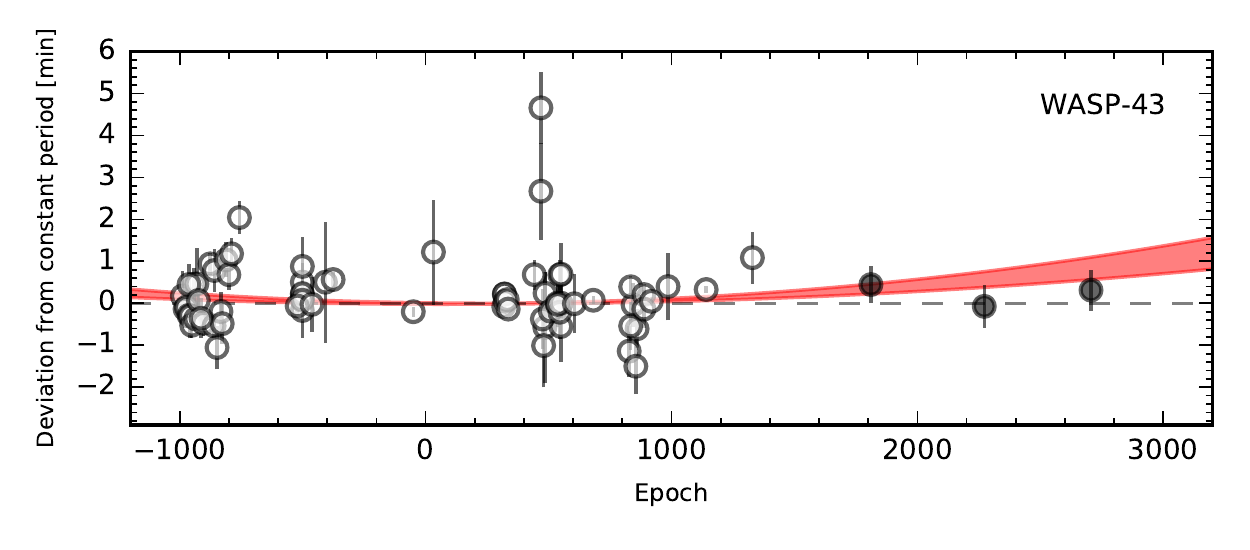}
	\caption{Timing residuals for WASP-43\,b. White points are
          times adopted from the literature: \cite{Gillon+2012,
            Chen+2014, Murgas+2014, Stevenson+2014, Ricci+2015,
            Jiang+2016, Hoyer+2016, Stevenson+2017}. Black points show new transits and the red band
          represents the 1-$\sigma$ uncertainty in the orbital decay
          model.}
	\label{fig:wasp43oc}
\end{figure*}

\item \textbf{WASP-114\,b} is a 1.8~$M_{\rm Jup}$ planet in 1.5-day orbit
  around a star somewhat hotter and more massive than the Sun. It was discovered recently, by \cite{Barros+2016}.
  The single new transit observed at FLWO is the only follow-up
  observation that has been reported. Since there are only two points in
  the timing residuals plot, it does not make much sense to fit an orbital decay model.
  The red band in Figure \ref{fig:wasp114oc}, therefore
  represents the 1-$\sigma$ uncertainty on the linear ephemeris, which
  we calculate to be $t_{\rm tra}(E) = 2456667.73661(21) ~\rm{BJD_{TDB}}$ + $E \times 1.548777461(81)~ \rm{ days}$.

\begin{figure*}
	\centering
	\includegraphics[width=1.0\linewidth]{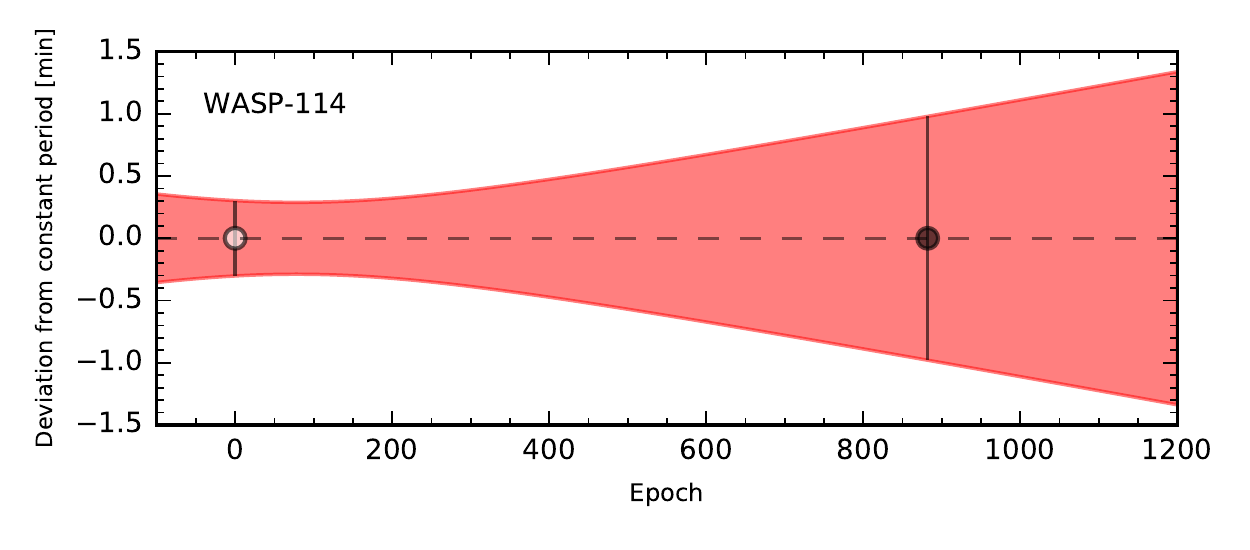}
	\caption{Timing residuals for WASP-114\,b. The white point is from
          \citet{Barros+2016}. and the black point represents new
          data point. The red band depicts 1$\sigma$ uncertainty on the
          linear ephemeris.}
	\label{fig:wasp114oc}
\end{figure*}

\item \textbf{WASP-122\,b} is a 1.3~$M_{\rm Jup}$ planet in a 1.7-day
  orbit around a G star. It was discovered by
  \cite{Turner+2016}. Unfortunately, we were not able to obtain any
  new data, so we present it here simply to call attention to its
  favorable properties and its place in the ``top dozen'' systems to
  watch for orbital decay.

\end{enumerate}

\section{Summary}
\label{sec:summary}

\begin{table*}
	\centering
	\begin{tabular}{l|cc|ccc|c}
		\hline 
		\hline
		& constant period && orbital decay & & &  \\
		Name & $t_{0}$ (BJD\textsubscript{TDB}) & $P$ (d) &  $t_{0}$ (BJD\textsubscript{TDB}) & $P$ (d) & $ dP/dt$ & $Q_{\star}^{\prime}$ \\
		\hline
		WASP-18    &2458022.12523(02) & 0.941452425(22) &2458022.12529(06)  &0.94145230(13)   & $(-1.2\pm 1.3) \times 10^{-10}$   & $>\,(1.7\pm 0.4) \times 10^6$ \\ 
		KELT-16    &2457910.03913(11) & 0.96899319(30)  &2457910.03918(15)  &0.96899314(33)   & $(-0.6\pm 1.4) \times 10^{-9}$    & $>\,(0.9\pm 0.2) \times 10^5$  \\ 
		WASP-103   &2456836.29630(07) & 0.925545352(94) &2456836.29635(07)  &0.92554483(25)   & $( 8.4\pm 4.0) \times 10^{-10}$   & $>\,(1.1\pm 0.1) \times 10^5$ \\
		WASP-12    &2456305.45556(03) & 1.091419810(39) &2456305.45581(04)  &1.091420087(46)  & $(-9.97\pm 0.83)\times 10^{-10}$  & $=\,(1.6\pm 0.2) \times 10^5$ \\
		HATS-18    &2457089.90665(25) & 0.83784309(28)  & \nodata & \nodata & \nodata  & \nodata    \\ 
		WASP-19    &2456021.70390(02) & 0.788839300(17) &2456021.70395(02)  & 0.788839420(31) & $(-2.06\pm 0.42) \times 10^{-10}$ & $=\,(5.0\pm 1.5) \times 10^5$ \\
		OGLE-TR-56 &2453936.60063(14) & 1.21191123(16)  &2453936.60059(14)  &1.21191089(23)   & $(3.34\pm 1.49) \times 10^{-10}$  & $>\,(7.0\pm 2.1) \times 10^5 $ \\ 
		HAT-P-23   &2456129.43472(09) & 1.212886411(72) &2456129.43466(11)  &1.21288633(11)   & $(1.47\pm 1.48) \times 10^{-10}$  & $>\,(6.4\pm 1.9) \times 10^5$ \\ 
		WASP-72    &2455583.6552(11)  & 2.2167360(27)   &2455583.6542(20)   &2.216742(11)     & $(-5.7\pm 9.2)\times 10^{-9} $    & $>\,(2.1\pm 1.4) \times 10^3$ \\
		WASP-43    &2456342.34259(01) & 0.813474059(29) &2456342.34259(01)  &0.813474024(31)  & $(1.9\pm 0.6) \times 10^{-10}$    & $>\,(4.4\pm 0.3) \times 10^5$\\ 
		WASP-114   &2456667.73661(21) & 1.548777461(81) & \nodata & \nodata & \nodata & \nodata \\ 
		WASP-122   &2456665.22401(21) & 1.7100566(29)   & \nodata & \nodata & \nodata & \nodata \\ 
		\hline 
	\end{tabular} 
	\caption{Summary of best-fit parameters and $Q_{\star}^{\prime}$. The lower limits on $Q'_\star$ are based on the 95\%-confidence
          lower limits on $dP/dt$, and the quoted uncertainties come from propagating the errors in $M_{\rm p}/M_\star$ and $a/R_\star$.}
	\label{tbl:summary}
\end{table*}

In this paper, we have drawn up a list of exoplanets that provide the
best opportunities to detect orbital decay, according to a simple
metric. We presented new transit times for 11 of those
exoplanets. Except for the case of WASP-12, we do not see convincing
evidence for a period change in any other system. There is some
evidence for a decreasing period in WASP-19, but the data are sparse
and there are some deviant data points, along with some indications
that stellar activity is causing systematic errors in the transit
times.  This is nevertheless a good system to continue monitoring
because of its extremely short period. The non-detections of orbital
decay allows us to constrain the tidal quality factor for the star,
the strongest limit being obtained for WASP-18: $Q_{\star}^{\prime} >
(1.7\pm 0.4) \times 10^{6}$ with 95\% confidence. Table~\ref{tbl:summary}
provides a summary of the best-fit parameters for all objects along
with limits on $Q_{\star}^{\prime}$.

This is a unique time for studying orbital decay of exoplanets. The
{\it TESS} mission is now underway. The main mission is to perform a
nearly all-sky survey for new transiting planets. \cite{Christ+2018}
discussed the prospects for {\it TESS} in detecting orbital decay of
{\it Kepler} planets, in particular, finding that there are at least a
few systems that offer good prospects. In the course of the survey,
{\it TESS} will also revisit almost all of the hot Jupiters that were
previously discovered in wide-field surveys, including the dozen
objects highlighted in this paper.  This may lead to additional
detections of period changes, or at the very least a helpful
refinement in the transit ephemerides.

\acknowledgements We are very grateful to Allyson Bieryla, David Latham,
and Emilio Falco for their assistance with the FLWO
observations.

This research has made use of The Extrasolar Planets Encyclopedia at
\textit{exoplanet.eu}. We also acknowledge the use of the NASA
Exoplanet Archive, which is operated by the California Institute of
Technology, under contract with the National Aeronautics and Space
Administration under the Exoplanet Exploration Program. The research
leading to these results has received funding from the ARC grant for
Concerted Research Actions, financed by the Wallonia-Brussels
Federation. TRAPPIST is funded by the Belgian Fund for Scientific
Research (Fond de la Recherche Scientifique, F.R.S-FNRS) under the
grant FRFC 2.5.594.09.F, with the participation of the Swiss National
Science Fundation (SNF). MG and EJ are F.R.S.-FNRS Senior Research
Associates. This work is partly supported by JSPS KAKENHI Grant
Numbers JP18H01265 and 18H05439, and JST PRESTO Grant Number
JPMJPR1775.  LD acknowledges support from the Gruber Foundation
Fellowship.  Work by JNW was supported by the Heising-Simons
Foundation.  TRAPPIST-North is a project funded by the University of
Li\`ege, and performed in collaboration with Cadi Ayyad University of
Marrakesh.

\bibliographystyle{yahapj}
\bibliography{references}

\newpage

\end{document}